\documentclass[preprint, 5p, twocolumn]{elsarticle}
\journal{Physics Letters A}
\usepackage{epsfig,newlfont,bm,subfigure,palatino,mathtools,braket,times,soul,setstack}
\usepackage[utf8]{inputenc}
\usepackage[sc,osf]{mathpazo}
\usepackage{amsmath}
\usepackage[T1]{fontenc}
\usepackage{latexsym}
\usepackage{amssymb}
\usepackage[colorlinks=true,citecolor=blue,urlcolor=blue]{hyperref}
\usepackage{color}
\usepackage{graphics,epstopdf}
\usepackage{soul}
\usepackage{graphicx}
\usepackage{capt-of}
\usepackage{lipsum}
\usepackage{adjustbox}
\usepackage[normalem]{ulem}
\usepackage[table,xcdraw]{xcolor}
\usepackage{braket}
\usepackage{physics}
\usepackage{ragged2e}
\usepackage{mathtools}

\usepackage{comment}
\usepackage{ragged2e}
\usepackage{braket}
\usepackage[justification=justified]{caption}
\usepackage[font=small,labelfont=bf,justification=justified,singlelinecheck=false]{caption}
\definecolor{indiagreen}{rgb}{0.07, 0.53, 0.03}
\definecolor{teal}{rgb}{0.0, 0.53, 0.53}
\newcommand{\tm}[1]{{\color{black} #1}}

\begin{document}

\title{Security of two-way deterministic quantum key distribution with higher-dimensional systems}

\author{ Abhishek Muhuri$^{1,2}$, Ayan Patra$^1$, Rivu Gupta$^3$, Tamoghna Das$^4$, Aditi Sen(De)$^1$}

\address{$^1$ Harish-Chandra Research Institute,  A CI of Homi Bhabha National Institute, Chhatnag Road, Jhunsi, Prayagraj - $211019$, India}
\address{$^2$ Center for Quantum Science and Technology, International Institute of Information Technology Hyderabad, Gachibowli, Hyderabad, Telangana 500032, India}
\address{$^3$ Dipartimento di Fisica “Aldo Pontremoli,” Università degli Studi di Milano, I-$20133$ Milano, Italy}
\address{$^4$Department of Physics, Indian Institute of Technology Kharagpur, Kharagpur 721302, India}

\begin{abstract}


We analyze the security of two-way quantum key distribution using arbitrary finite-dimensional systems, considering both individual and collective eavesdropping attacks, without the effective use of entangled states,
 by incorporating two mutually unbiased bases and Heisenberg-Weyl operators in higher dimensions. For individual attacks, we consider cloning operations by the eavesdropper and demonstrate a dimensional advantage where secret keys can be generated for greater strengths of interception. To analyze security under collective attacks, we employ a purification scheme and derive the key rate using entropic uncertainty relations. Further, we exhibit how the protocol is more robust against eavesdropping with increasing dimension of the systems used, and compare the performance with that of the entangled two-way secure dense coding protocol when the presence of the eavesdropper is modeled by correlated and uncorrelated noise.

\end{abstract}

\maketitle

\section{Introduction}

Quantum key distribution (QKD)~\cite{Gisin_RMP_2002_cryptography-review, Scarani_RMP_2009_cryptography-review} is a fundamental information-theoretic protocol ensuring security while transmitting information between distant parties. The protocol leverages the inherent non-classical properties of quantum mechanics, such as entanglement~\cite{Horodecki_RMP_2009_entanglement-review}, nonlocal correlations between the honest parties \cite{Bell-nonlocality, Ekert_PRL_1991_Ekert-protocol,  Mayers_IEEE_1998_device-independent-security, Gottessman_IEEE_2004_device-independent-security, Acin_PRA_2006_device-dependent-BBBMM, Acin_PRL_2007_device-independent-security, Masanes_Nature-Comm_2011_device-independent-security, Pironio_PRX_2013_device-independent-security, Vazirani_PRL_2014_device-independent-security, Miller_JACM_2016_device-independent-security}, and 
non-orthogonal states ~\cite{Nielsen_2010_book, Wilde_2013_book,  Preskill_2015_book, Watrous_CUP_2018_book} to ensure that the malicious parties can never intercept the entire information without being detected. Different protocols exist for generating secret keys, 
viz., the BB$84$~\cite{Bennett_TCS_2014_BB84}, Ekert$91$ ~\cite{Ekert_PRL_1991_Ekert-protocol}, and the six-state protocols~\cite{Bruss_PRL_1998_six-state, Bechmann_PRA_1999_six-state}, 
which are essentially one-way schemes involving the transfer of quantum bits between the honest parties. However, such protocols are inherently probabilistic~\cite{Tomamichel_NC_2012_tight-finite-key} since they require discarding several qubits if basis reconciliation is unsuccessful. To circumvent this issue, deterministic key distribution protocols have also been introduced, such as the Ping-Pong~\cite{Bostrom_PRL_2002_ping-pong} and LM$05$ protocols~\cite{Lucamracini_PRL_2005_LM05}, which, however, require two-way quantum communication from the sender to the receiver and back. The security of these protocols has also been well established~\cite{Cai_PRL_2003_Ping-pong-insecure, Wojcik_PRL_2003_Ping-pong-insecure, Deng_PRA_2003_Ping-pong-security, Deng_PRA_2004_Ping-pong-security, Cai_PRA_2004_improved-ping-pong, Lu_PRA_2011_LM05-secure, Fung_PRA_2012_LM05-secure}, although they allow the eavesdropper, Eve, to intercept the messages at two different stages of the communication. 

While a majority of cryptographic schemes have been proposed and analyzed using qubits, recent studies have focused on establishing the security of one-way QKD protocols by employing higher-dimensional systems or qudits~\cite{Bechmann_PRA_2000,Bourennane_PRA_2001_high-d-QKD, Bourennane_JPA_2002_high-d-QKD-security,  Nikolopoulos_PRA_2006, Sasaki_Nature_2014, Xavier_CP_2020_high-d-QKD-optical-fibre, Cerf_PRL_2002_high-d-BB84-six-state,Ogrodnik_arXiv_2024_high-d-QKD-resource-efficient}, using mutually unbiased bases (MUBs)~\cite{Wyderka_arXiv_high-d-QKD-MUB}, as well as Fourier-bases~\cite{Scarfe_arXiv_high-d-QKD-Fourier-basis}, both in the asymptotic
\cite{Thomas_arXiv_high-d-QKD-asymptotic-key} and  realizable finite-size limits~\cite{Thomas_arXiv_high-d-QKD-finite-key}. 
It has also been widely recognized that the performance of qudit-based quantum devices is superior to that of qubit systems~\cite{Correa_PRE_2014, Wang_PRE_2015, Santos_PRE_2019, Dou_EPL_2020, Usui_PRA_2021, Ghosh_PRA_2022, Konar_PRA_2023,Wei_PRL_2019, Nagali_PRL_2010, Bouchard_SA_2017,Ecker_PRX_2019,Lanyon_NP_2009, Babazadeh_PRL_2017, Muralidharan_NJP_2017}. 
Further, qudits can be experimentally generated through twisted photons~\cite{Bouchard_Quantum_2018}, silicon-integrated circuits~\cite{Ding_NPJ_2017}, microwave~\cite{Sit_OL_2018, Amitonova_OE_2020}, telecommunication fibres~\cite{Canas_PRA_2017}, and time-bin encoding~\cite{Islam_SA_2017}, ensuring realization of dimensional benefits in laboratories. 

\begin{figure}[t]
\includegraphics[width=1.0\linewidth]{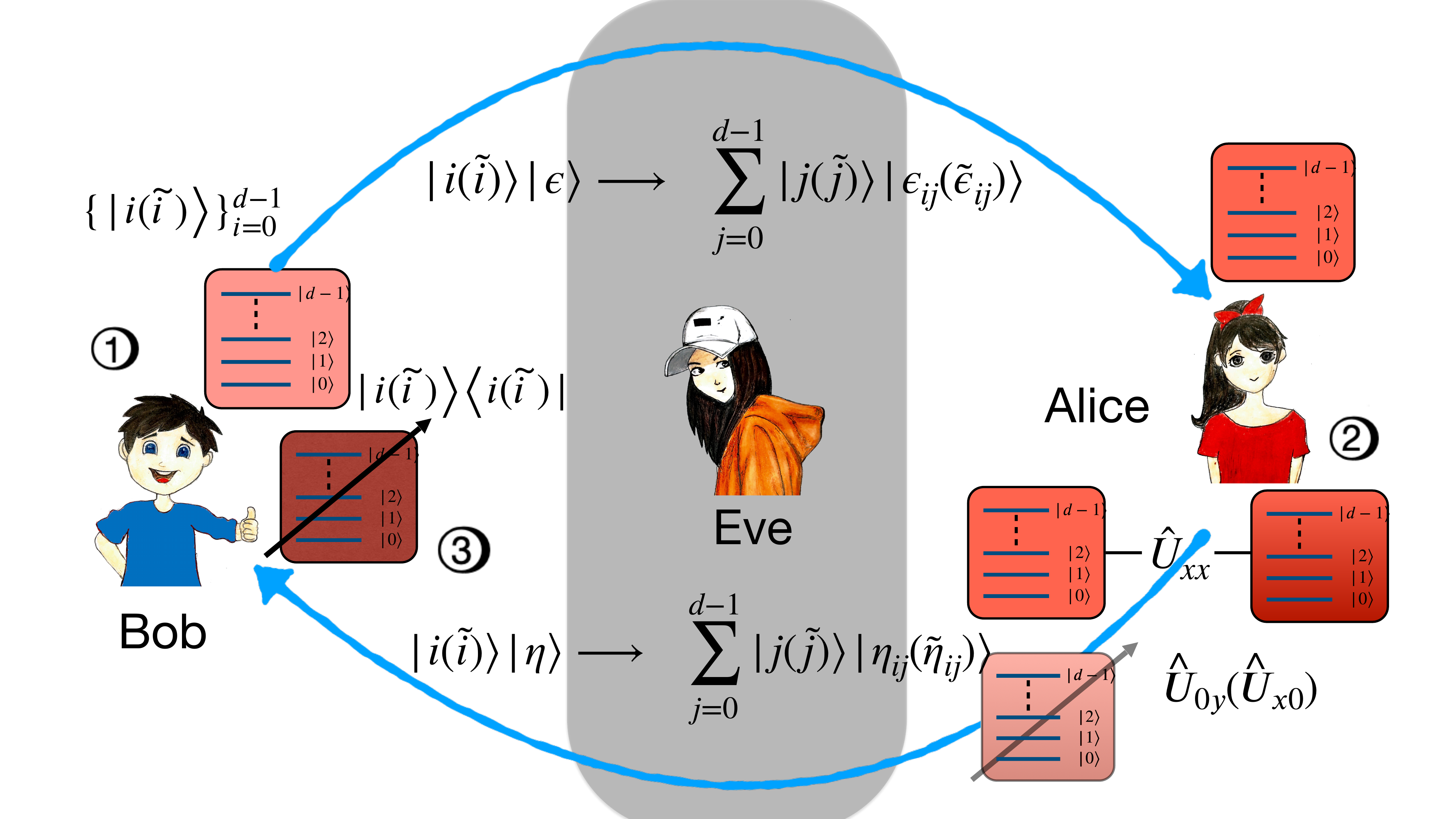}
\caption{\justifying {Schematic diagram of the higher-dimensional, entanglement-free, and two-way quantum key distribution protocol under a generalized individual attack scenario, using qudits. Two quantum channels are established between Alice and Bob: the first channel allows Bob to send the prepared state $\ket{i(\tilde{i})}_B$ to Alice, while the second one enables Alice to return the encoded state to Bob. Alice’s encoding operations are represented by unitary operators, $\hat{U}_{xx}$, where $x,y \in \{0,1\}$ are the encoding choices of Alice, whereas $\hat{U}_{0y} (\hat{U}_{x0})$ comprises the operations for the test runs. Eve may intercept and attack both quantum channels independently using a probabilistic quantum cloning machine.
}}
\label{fig:schematic} 
\end{figure}

It is thus an important problem to address the performance of two-way cryptographic protocols using higher-dimensional systems.
In this paper, we design an entanglement-free two-way QKD protocol, \tm{a generalization of} the LM$05$ scheme, in {\it arbitrarily large finite dimensions } (as shown in Fig. \ref{fig:schematic}), addressing the gap in the existing literature that considered only prime-power-dimensional systems~\cite{Eusebi_QIC_2009_high-d-LM05} using the corresponding MUBs~\cite{Ivonovic_JPA_1981_high-d-MUB, Wootters_AP_1989_high-d-MUB, Bandopadhyay_Algorithmica_2002_high-d-MUB, Mullen_2004_book}.
Further, we demonstrate the \tm{security of the proposed higher dimensional protocol} against both individual cloning attacks and the more powerful collective attacks by an eavesdropper, \tm{who is bounded only by the law of quantum mechanics}. In the individual attack \tm{model, we consider that} each \tm{transmitted} qudit \tm{from Bob to Alice and then back to Bob} \tm{are both} intercepted \tm{by the eavesdropper} and measured separately ~\cite{Beaudry_PRA_2013_two-way-QKD}, whereas \tm{in case of} collective attacks, \tm{we} assume that the eavesdropper possesses quantum memory, allowing joint measurements \tm{on her part of the quantum system,} after accumulating quantum \tm{and classical post processing} data (see Ref.~\cite{Beaudry_PRA_2013_two-way-QKD} for the qubit case).
In the former situation, we consider a general cloning strategy based on minimal {possible} assumptions and explicitly elucidate how a non-vanishing secret key rate is attained because the information shared between the honest parties is greater than that shared with Eve, {when the latter has the choice of using a cloning machine with minimal detection probability}.
thereby leading to a non-vanishing secret key rate. 

For collective attacks by the eavesdropper, we {\it introduce} a modified version of the protocol that incorporates a purification of the preparation and encoding steps, enabling us to establish its security against this attack. 
\tm{In this case, the presence of Eve can be incorporated in the system by giving her the additional interface of the purification of the quantum state shared by the honest parties. The protocol considers}
the influence of a generic \tm{eavesdropping} attack modeled by different noisy channels. \tm{This model of the eavesdropper eventually gives ultimate operational power to gain maximum information about the honest parties as much as possible \cite{Shor_PRL_2000_device-dependent-BB84,Kraus_PRL_2005_device-dependent-security}}.  We illustrate how one can obtain secret keys for qudits under greater levels of attack, \tm{i.e., the worst case scenario for Alice and Bob, where the purifying system held by Eve allows complete access to the information of their quantum system}. Hence, security proof against collective attack does not  \tm{require} any specific attack strategy, unlike in the case of individual attacks. In both \tm{the} situations of \tm{individual and collective attacks}, we report the {\it dimensional advantage} through an increase in the key rate with increasing dimension. 

{\it Hierarchy between securities in two-way entanglement-free and entanglement-based protocols.} In the regime of two-way QKDs,  prominent examples of entanglement-free and entanglement-based protocols include the LM$05$ and the secure dense coding (SDC)~\cite{Bostrom_PRL_2002_ping-pong, Lucamracini_PRL_2005_LM05, Beaudry_PRA_2013_two-way-QKD} schemes. 
A key distinction between these two protocols lies in the number of secret key bits generated in each consecutive run of the protocol - the SDC protocol, relying on the dense coding scheme, produces two bits of secret key, \tm{because of its pre-shared one ebit of entanglement between the honest parties}. In contrast, the LM$05$ protocol guarantees the security of only one bit in the noiseless limit, \tm{without any entanglement present in the protocol}. In a recent work~\cite{Patra_PRA_2024_dimensional-advantage-SDC}, some of us have shown that qudits can provide higher secret key rates than qubits even in the two-way  SDC protocol~\cite{Bostrom_PRL_2002_ping-pong, Beaudry_PRA_2013_two-way-QKD}. This work presents a fair comparison between the two, utilizing qudits of larger dimensions or a greater number of runs in the LM$05$ case, where the presence of the malicious eavesdropper is modeled through independent and correlated noise channels.
Juxtaposing the performance of the two protocols in the presence of noise would highlight whether entanglement provides additional robustness against adversarial attacks or if the cumbersome generation of high-dimensional entanglement can be circumvented at the expense of preparing qudits with larger dimensions. 
Depending upon the resources available or the ease of producing a particular kind of state, whether entangled or non-orthogonal, the results of our study can be used to choose the protocol best suited for two-way key distribution using qudits. Specifically, we show that the qudit-based LM$05$ protocol yields a higher secure key rate than that of the SDC  when subjected to uncorrelated noise. In contrast, under correlated noise, the SDC protocol outperforms the LM$05$.

The article is organized in the following way. In Sec.~\ref{sec:individual_attack}, we define the steps comprising the high-dimensional LM$05$ protocol and establish dimensional advantage in the security in terms of the probability of detecting an eavesdropper performing a two-way individual cloning attack. We provide a rigorous derivation of the key rate through exact expressions of the mutual information between the different parties. We then proceed to analyze the efficacy of the protocol under collective attacks in Sec.~\ref{sec:collective_attack}, where we consider a modified version of the same involving purification and entropic uncertainty relations. We include concluding discussions in Sec.~\ref{sec:conclu}.

\section{Security of the high-dimensional LM$\mathbf{05}$ protocol against individual attacks}
\label{sec:individual_attack}

In this section, we establish how, using higher-dimensional systems, the LM$05$ protocol can tolerate higher levels of interference from the eavsdropper, as compared to qubit version~\cite{Lucamracini_PRL_2005_LM05}, thereby providing dimensional benefit in terms of security. Let us first describe the protocol in arbitrary finite dimensions, while simultaneously recapitulating the same for qubits.

\subsubsection*{The high-dimensional LM$05$ key-distribution protocol}

The protocol consists of two honest parties, say Alice ($A$) and Bob ($B$), while we denote the eavesdropper as Eve ($E$). It proceeds in three main steps - the preparation stage, the encoding stage, and the measurement stage.

{\bf Preparation:} $B$ randomly prepares a state, $\rho_B$, either from the computational basis set, $\{\ket{i}\}$, or the Fourier basis, $\{\ket{\tilde{i}} = \frac{1}{\sqrt{d}} \sum_{k = 0}^{d - 1} \omega^{ik} \ket{k}\}$, where $i \in [0, d-1]$ and $\omega = e^{2 \pi \iota/d}$ is the $d$-th root of unity (with $\iota = \sqrt{-1}$)
For $d = 2$, this corresponds to the eigenbases of either  Pauli-$Z$ or  Pauli-$X$ operators. Subsequently, $B$ stores $i$ as his input dit.

{\bf Encoding:} The party $A$, upon receipt of the state $\ket{i(\tilde{i})}_B$ through a quantum channel, $\mathcal{E}_1^{B \to A}$ (which henceforth we refer to as the {\it forward channel}), either encodes the message with an overwhelmingly high probability, $c \to 1$, or performs a security check with probability $(1-c)$. 
\begin{itemize}
    \item {\it Message mode with probability $c$.} Defining $\hat{U}_{xy} = \sum_{l = 0}^{d - 1} \omega^{ly} |l \oplus x\rangle \langle l| \footnote{Throughout the manuscript, $\oplus$ denotes summation modulo $d$.}, ~ \text{with}~ \{x, y\} \in [0, d-1]$ as the Heisenberg-Weyl operator  for $d \geq 2$~\cite{Hiroshima_JPA_2001_dense-coding}, the message mode entails the operation of $\hat{U}_{xx}$ on the received qudit, i.e., $A$ prepares $\rho'^{xx}_A = \hat{U}_{xx} \mathcal{E}_1^{B \to A} (\rho_B) \hat{U}_{xx}^\dagger$.  For qubits, $A$ applies either the identity $(\hat{U}_{00})$ or the Pauli-$Y$ $(\hat{U}_{11})$ gate for encoding. The value of $x$ is then recorded as the key \textcolor{black}{and the encoded state $\rho'^{xx}_A$ is then sent back to $B$ through the {\it backward channel} $\mathcal{E}_2^{A \to B}$.}
    \item {\it Check mode with probability $1 - c$.} $A$ measures either on the $\hat{U}_{0y}$ or $\hat{U}_{x0}$ basis (Pauli-$Z$ or the Pauli-$X$ basis for $d = 2$). The modified system is then sent back to $B$ through the channel, $\mathcal{E}_2^{A \to B}$. 
\end{itemize}
{\bf Measurement:} The final step involves $B$ measuring $\rho'^{xx}_B = \mathcal{E}_2^{A \to B} (\rho'^{xx}_A)$ in the same basis in which $\rho_B$ was initially prepared. Upon measurement, the outcome, say ``$o$'', is stored as the output dit.\\

\noindent \textit{Post-processing}. After repeating the above steps $N$ times, $A$ reveals which rounds were measured (and in which basis), and which were encoded, i.e., which rounds were used as message modes. In the encoding runs, i.e., the message mode, $B$ computes the key value by performing a generalized XOR operation (summation modulo $d$) between $o$ and $i^C = d - i$, while $A$ possesses the key value $x$, corresponding to her encoding operation $\hat{U}_{xx}$. In the check rounds, $A$ and $B$ discard their outcomes whenever their measurement bases do not coincide; otherwise, they can detect the presence of Eve by comparing their measurement outcomes. In particular, the absence of a perfect double-correlation between the measurement bases on the forward and backward channels, during the check runs, signals the presence of $E$.


Note that the protocol is deterministic, as no key dits are discarded due to basis mismatch during post-processing. Furthermore, the absence of classical communication about the encoding operation enhances the security of the protocol~\cite{Lucamracini_PRL_2005_LM05}. We now proceed to analyze the protocol's security under cloning attacks.

\subsection{Dimensional benefit in two-way individual cloning attack}
\label{subsec:cloning_attack}

Since LM$05$ is a two-way protocol comprising two channels, the eavesdropper can attack twice in a bid to attain as much information as possible without being detected~\cite{Beaudry_PRA_2013_two-way-QKD}. Let us describe the eavesdropper's strategy when limited to interfering with each transmitted qudit separately, explain the protocol when affected by such an attack, and finally derive the key rate. One of the prominent individual attack strategies is to append auxiliary \textcolor{black}{systems} to the original system, perform a cloning operation, and, by measuring the auxiliary subsystem, try to decipher the secret message~\cite{Lucamracini_PRL_2005_LM05}. We can define this {\it cloning attack} by the unitary transformation,

\begin{eqnarray}
    \ket{i (\tilde{i})}_{A/B} \ket{\epsilon}_E \to \sum_{j = 0}^{d - 1} \ket{j (\tilde{j})}_{A/B} \ket{\epsilon_{ij} (\tilde{\epsilon}_{ij})}_E,
    \label{eq:basis_clone}
\end{eqnarray}
where $\ket{\tilde{\epsilon}_{ij}} = \frac{1}{d} \sum_{k,l = 0}^{d - 1} \omega^{ik - lj} \ket{\epsilon_{kl}}$ follows from the linearity of the cloning transformation on the states $\ket{\tilde{i}}$. 
Let us assume that for the computational basis states, $\ket{i}$, Eve possess the state $\ket{\epsilon_{ii}}$, when the attacked qudit is cloned perfectly, whereas $\ket{\epsilon_{ij}}$ corresponds to the latter being modified from $\ket{i} \to \ket{j}$ (and similarly for the basis $\{\ket{\tilde{i}}\}$). For the moment, we do not consider normalized auxiliary states. 

\textcolor{black}{Let us consider that the cloning machine treats all computational basis states uniformly, such that each state $\ket{i}$ is cloned with probability $F$, and is flipped to a different state $\ket{j}$ (where $j \neq i$) with probability $D = (1-F)/(d-1)$, for all $i,j$. Therefore, we can write the overlaps as $\langle \epsilon_{ii} | \epsilon_{ii} \rangle = F~\forall ~i$, and $\langle \epsilon_{ij} | \epsilon_{ij} \rangle = D$ for $i \neq j$. The unitarity of the cloning transformation imposes the following conditions:
\begin{itemize}
   \item[\textbf{CI.}] $F + (d-1)D = 1$,
   \item[\textbf{CII.}] $\sum_k \langle \epsilon_{ik} | \epsilon_{jk} \rangle = 0 \quad \forall~i,j$.
\end{itemize}
It is important to note that, since the protocol does not involve basis reconciliation,  an eavesdropper may exploit this to perform an asymmetric cloning attack. At this stage, we introduce a minimal set of assumptions regarding the cloning transformation, at the cost of a small loss of generality (otherwise, the mathematical treatment would be intractable), to analyze the security of the protocol. 
\begin{enumerate}
    \item[\textbf{AI.}] For a fixed input state $\ket{k}$, the corresponding states of Eve after the cloning operation are mutually orthogonal, i.e., $\langle\epsilon_{ki}|\epsilon_{kj}\rangle=0~\forall~k,$ with $i\neq j$.
    \item[\textbf{AII.}]  For a fixed output index $k$, the overlap between Eve's states corresponding to different input states is zero, $\langle\epsilon_{ik}|\epsilon_{jk}\rangle=0~\forall~k,$ with $i\neq j$. This assumption guarantees that condition \textbf{CII} is satisfied, albeit at the expense of some generality.
    \item[\textbf{AIII.}] Finally, we assume that Eve's states corresponding to the perfect cloning, i.e., $\{\ket{\epsilon_{ii}}\}_{i=0}^{d-1}$, span a subspace that is orthogonal to the subspace spanned by the states corresponding to incorrect cloning (flipping), i.e., $\{\ket{\epsilon_{ij}}_{i\neq j}\}_{i,j=0}^{d-1}$. Mathematically, this means $\{\ket{\epsilon_{ii}}\}_{i=0}^{d-1}\perp \{\ket{\epsilon_{ij}}_{i\neq j}\}_{i,j=0}^{d-1}$, and  $\{\ket{\epsilon_{ii}}\}_{i=0}^{d-1}\oplus \{\ket{\epsilon_{ij}}_{i\neq j}\}_{i,j=0}^{d-1}=\mathcal{H}_E$, where $\mathcal{H}_E$ denotes Eve's Hilbert space.
\end{enumerate}
The overlaps between all relevant pairs of Eve's states resulting from the cloning process read
\begin{align}
\label{eq:eij_norm}
    \langle \epsilon_{ij}| \epsilon_{ij}\rangle &= \begin{cases}
         F ~ \text{for} ~ j = i \\
         \frac{1 - F}{d - 1} ~ \text{for} ~ j \neq i~,
    \end{cases}\\
\label{eq:eii_ejj_innerproduct}  
    \langle \epsilon_{ii}| \epsilon_{jj}\rangle &= F \cos \theta_{ij}~, \\ 
\label{eq:eij_ekl_innerproduct} 
    \langle \epsilon_{ij}| \epsilon_{kl}\rangle &= \frac{1 - F}{d - 1} \cos \phi_{ijkl} \text{ for } i\neq k, j\neq l,\text{ and}\notag\\
    &\phantom{= \frac{1 - F}{d - 1} \cos \chi_{ijkl} \text{ for }} ~i\neq j, k\neq l,
\end{align}
where $0\leq\{\theta_{ij},\phi_{ijkl}\}\leq\pi/2$, $\theta_{ij}=\theta_{ji}$, and $\phi_{ijkl}=\phi_{klij}$. In accordance with our previous assumptions, the remaining auxiliary states are mutually orthogonal.
}
\textcolor{black}{
Further analysis requires evaluating $\langle \tilde{\epsilon}_{ii}| \tilde{\epsilon}_{ii}\rangle$, i.e., the probability of perfectly cloning the state $\ket{\tilde{i}}$, which can be obtained from Eqs.~\eqref{eq:eij_norm} - \eqref{eq:eij_ekl_innerproduct} as
\begin{align}
    \langle \tilde{\epsilon}_{ii}| \tilde{\epsilon}_{ii} \rangle = \frac{1}{d^2} \Bigg(  \, d &+ F \sum_{\substack{j,k \\ j \neq k}} \cos \theta_{jk} \nonumber \\
    & + \frac{1 - F}{d - 1} \sum_{\substack{j,k,l,m \\ j \neq k, l \neq j \\ m \neq k, l \neq m}} \omega^{i(l - m + k - j)} \cos \phi_{jklm} \Bigg). \label{eq:e-tilde_orthogonality}
\end{align}
Similar relations hold for $\ket{\eta_{ij}}$, the auxiliary states used for the attack on the backward channel, by substituting $F \to F', \theta_{ij} \to \theta'_{ij}, ~ \text{and} ~ \phi_{ijkl} \to \phi'_{ijkl}$. 
}

\subsubsection{Probability of detecting the eavesdropper}
\label{subsubsec:detection_prob}

While we have derived the norms of the auxiliary states in the possession of $E$ for the forward channel, similar relations hold for $\ket{\eta_{ij}}$, the auxiliary states used for the attack on the backward channel, by substituting $F \to F', \theta_{ij} \to \theta'_{ij}, ~ \text{and} ~ \phi_{ijkl} \to \phi'_{ijkl}$. The presence of $E$ is unnoticed if the cloning transformation succeeds, i.e., the {\it non-detection} probabilities for $E$ are given by $P_{\text{nd}}^{B \to A}(\ket{i (\tilde{i})}) = \langle \epsilon_{ii} (\tilde{\epsilon}_{ii}) | \epsilon_{ii} (\tilde{\epsilon}_{ii})\rangle$ for the forward channel, as can be obtained from Eq.~\eqref{eq:e-tilde_orthogonality}, while similar expressions are valid for the backward one with the aforementioned substitutions. Thus, on average, the eavesdropper can be detected with probability

\begin{equation}
    \bar{P}_{\text{det}} = 1 - \frac{1}{2d} \sum_{\ket{\psi}\in \{\ket{i},\ket{\tilde{i}}\}_{i=0}^{d-1}} P_{\text{nd}}^{B \to A} (\ket{\psi})\times P_{\text{nd}}^{A \to B}(\ket{\psi}),
    \label{eq:detection_prob}
\end{equation}
where the summation is over the $2d$ initial states comprising the two bases. \textcolor{black}{From Eqs.~\eqref{eq:eij_norm}, \eqref{eq:e-tilde_orthogonality}, and \eqref{eq:detection_prob}, it follows that $\bar{P}_{\text{det}}$ is a sum of bi-linear and affine functions of the variables $F$ and $F'$, i.e., $\bar{P}_{\text{det}}=a F F'+b F+cF'+d$, where the coefficients are functions of $\{\theta_{ij},\phi_{ijkl}\}$. Consequently, within the domain $0 \leq \{F, F'\} \leq 1$, the minimum value of $\bar{P}_{\text{det}}$ must be attained at one of the four corner points, $(F, F') \in \{(0,0), (0,1), (1,0), (1,1)\}$. Interestingly, similar to the qubit case studied in Ref.~\cite{Lucamracini_PRL_2005_LM05}, the minimum average detection probability, denoted by $\bar{P}_{\text{det}}^{\min}$, is achieved at $F = F' = 1$\footnote{We have verified this claim numerically. Since the minimization involves only four discrete points, the numerical analysis can be performed for arbitrary dimension. In particular, one can express $\bar{P}_{\text{det}}=a F F'+b F+cF'+d$, where the coefficients satisfy $a+b+c<0$, $a+b<0$, $a+c<0$, and $a<0$, for all values of the angular variables appearing in Eqs.~\eqref{eq:eii_ejj_innerproduct} and \eqref{eq:eij_ekl_innerproduct}. It then follows that the minimum of $\bar{P}_{\text{det}}$ occurs at $F=F'=1$.}, which indicates the optimal strategy for $E$. At the limit $F = F' = 1$, the cloning attack exhibits two notable properties -- $(i)$ the cloning machine perfectly clones all the computational basis states,  thereby allowing $E$ to induce a vanishing disturbance when such states are employed in the protocol, and $(ii)$ the machine treats all the Fourier basis states identically (see Eq.~\eqref{eq:e-tilde_orthogonality} with $F = 1$), indicating symmetric behavior with respect to that basis.\\
\textbf{Note $\mathbf{1}$.} We have also performed a numerical analysis for qutrit systems without invoking the assumption \textbf{AIII}, and found that the minimum detection probability for Eve still occurs at $F = F' = 1$. This leads us to conjecture that the assumption \textbf{AIII} does not compromise generality. 
}



\subsubsection{The protocol including Eve's attack}
\label{subsubsec:I_AB}

We now explicitly describe the LM$05$ protocol under the aforementioned eavesdropping strategy. Let us consider that the qudit state is prepared as $\ket{\psi}_B = \sum_n C_n \ket{n}_B$, where $C_n=\delta_{in}$ for $\ket{i}$, and $C_n=\frac{1}{\sqrt{d}} {\omega}^{in}$ for $\ket{\tilde{i}}$. The encoding operation with $\hat{U}_{xx}$, while suffering the cloning attack by $E$ on both channels, reads
\begin{eqnarray}
   \hspace{-2.5em}&& \ket{\psi}_B \ket{\epsilon}_E \ket{\eta}_E \xrightarrow[]{\mathcal{E}_1^{B \to A}} \sum_n C_n \sum_{n'} \ket{n'}_A \ket{\epsilon_{nn'}}_E \ket{\eta}_E \label{eq:E_1_attack} \\
   \hspace{-2.5em} &&  \xrightarrow[]{\hat{U}_{xx}} \sum_n C_n \sum_{n'} \omega^{n' x} \ket{n' \oplus x}_A \ket{\epsilon_{nn'}}_E \ket{\eta}_E \label{eq:encoding} \\
   \hspace{-2.5em}&&  \xrightarrow[]{\mathcal{E}_2^{A \to B}} \sum_n C_n \sum_{n'} \omega^{n' x} \ket{\epsilon_{nn'}}_E \sum_{n''} \ket{n''}_B \ket{\eta_{(n' \oplus x) n''}}_E \label{eq:E_2_attack}. ~~~
\end{eqnarray}
The eavesdropper is thus in possession of the $d$ ensembles $\left\{ \ket{\epsilon_{nn'}} \ket{\eta_{{(n' \oplus x) n''}}}  \right\}_{n,n',n'' = 0}^{d-1}$ 
for $x = 0, 1, \dots, d - 1$, each with $d^3$ states. 
Observe that the $d$ ensembles are mutually disjoint. Therefore, identifying a specific state immediately determines its corresponding ensemble. Eve's task is now to find out the ensemble in order to infer the encoding operation, thereby allowing her to guess the message. The analysis is considerably simplified in the limit, $F = F' = 1$, corresponding to the minimum average detection probability, since the only states constituting the ensembles are $\left\{ \ket{\epsilon_{nn}} \ket{\eta_{(n \oplus x)(n \oplus x)}} \right\}$ (see Table.~\ref{tab:Eve_ensembles}).

\begin{table}[]
\resizebox{0.4\textwidth}{!}{%
\begin{minipage}{0.59\textwidth}
			\centering
\begin{tabular}{|c|c|c|c|}
\hline
\multicolumn{1}{|c|}{$x = 0$}                                 & \multicolumn{1}{c|}{$x = 1$}                      & \multicolumn{1}{c|}{$\dots$} & \multicolumn{1}{c|}{$x = d - 1$}                  \\ \hline
$\ket{\epsilon_{00}} \ket{\eta_{00}}$                         & $\ket{\epsilon_{00}} \ket{\eta_{11}}$             & $\dots$                      & $\ket{\epsilon_{00}} \ket{\eta_{(d - 1)(d - 1)}}$ \\ \hline
$\ket{\epsilon_{11}} \ket{\eta_{11}}$                         & $\ket{\epsilon_{11}} \ket{\eta_{22}}$             & $\dots$                      & $\ket{\epsilon_{11}} \ket{\eta_{00}}$             \\ \hline
$\vdots$                                                      & $\vdots$                                          & $\vdots$                     & $\vdots$                                          \\ \hline
$\ket{\epsilon_{(d - 1)(d - 1)}} \ket{\eta_{(d - 1)(d - 1)}}$ & $\ket{\epsilon_{(d - 1)(d - 1)}} \ket{\eta_{00}}$ & $\dots$                      & $\ket{\epsilon_{(d - 1)(d - 1)}} \ket{\eta_{(d - 2)(d - 2)}}$ \\ \hline
\end{tabular}

\vspace{0.5em}
\caption{\justifying{States constituting the ensembles with $E$ corresponding to encoding through $\hat{U}_{xx}$ by $A$ in the limit $F=F'=1$. Each column represents an ensemble, with a total of $d$ such columns.
Each column contains $d$ rows, corresponding to $d$ distinct states of an ensemble.}} 
\label{tab:Eve_ensembles}
\end{minipage}}
\end{table}

We are now in a position to estimate the lower bound on the key rate given by $r  = I_{AB} - \min [I_{AE}, I_{BE}]$~\cite{Csiszar_IEEE_1978_Csiszr-Korner-individual-key-rate, Gisin_RMP_2002_cryptography-review}, where the three terms represent the mutual Shannon information, \textcolor{black}{defined as $I_{\mathcal{MN}} = H(\mathcal{M}) + H(\mathcal{N}) - H(\mathcal{MN})$, with $H(\mathcal{M}) = - \sum_i p_i^\mathcal{M} \log_2 p_i^\mathcal{M}$~\cite{Nielsen_2010_book} being the Shannon entropy of the probability distribution, $\{p_i^\mathcal{M}\}$, and the calligraphic letters denote the respective parties.} 


\subsection{The key rate under individual cloning attack}
\label{subsec:individual_key}

Let us derive the expressions for the mutual information terms to arrive at the key rate for the $d$-dimensional LM$05$ protocol when affected by the cloning attack described so far. We will then show how a non-zero secret key rate exists at higher levels of interception by the eavesdropper with an increase in the dimension of the systems in use. Note that the analysis becomes mathematically intricate due to the presence of numerous free parameters. Therefore, we restrict our study to Eve's optimal strategy, i.e., $F = F' = 1$, which not only simplifies the analysis but also yields the optimized mutual Shannon information quantities, $I_{AB}$, $I_{AE}$, and $I_{BE}$. 

\subsubsection{Mutual information between the honest parties}
\label{subsubsec:I_AB}

To estimate $I_{AB}$, we need to consider the instances when, despite intervention by $E$, the encoded state is correctly received by $B$. The situation is trivial when the state is among $\{\ket{i}\}_{i = 0}^{d - 1}$ since $E$ always clones such states perfectly, and thus the state with $B$ is the same one as sent by $A$ post encoding. Therefore, given that the prepared state is $\ket{i}$, the probability that Alice's and Bob's dit values coincide is $P_{A = B}^{\ket{i}} = 1~\forall~i$, 
and \textcolor{black}{the mutual information in this case is given by $I_{AB}^{\ket{i}} =  \log_2 d$.}

On the other hand, when the state belongs to the Fourier basis, $\{\ket{\tilde{i}}\}$, $B$ receives the actual encoded state only if the cloning operations by $E$ on both channels produce the state corresponding to the uninterrupted message run. This is possible if $(a)$ the cloning is perfect in both channels, i.e., $E$ remains undetected, which occurs with probability $p_1^{\ket{\tilde{i}}} = P_{\text{nd}}^{B \to A}(\ket{\tilde{i}}) \times P_{\text{nd}}^{A \to B}(\ket{\tilde{i}})$ where $P_{\text{nd}}(\ket{\tilde{i}})$ is given by Eq.~\eqref{eq:e-tilde_orthogonality}; and also $(b)$ when the cloning is imperfect at both the channels. Consider that initially, $\ket{\tilde{i}}$ gets morphed to $\ket{\tilde{i'}}$ due to $E$'s attack on $\mathcal{E}_1^{B \to A}$ whereafter the encoding transforms it into $\hat{U}_{xx} \ket{\tilde{i'}}$. It is possible that imperfect cloning of $\hat{U}_{xx} \ket{\tilde{i'}}$ in $\mathcal{E}_2^{A \to B}$ results in $\ket{\tilde{i''}} = \hat{U}_{xx} \ket{\tilde{i}}$, in which case $B$ receives the intended state. With the total number of flips corresponding to erroneous cloning at both channels being $(d - 1)^2$ and for each of the $(d - 1)$ states $\hat{U}_{xx} \ket{\tilde{i'}}$, only one possible incorrect cloning operation which leads to $\hat{U}_{xx} \ket{\tilde{i}}$, this situation has a probability $ p_2^{\ket{\tilde{i}}} = (d - 1)^{-1} P_{\text{det}}^{B \to A}(\ket{\tilde{i}}) \times P_{\text{det}}^{A \to B}(\ket{\tilde{i}})$ where $P_{\text{det}} = 1 - P_{\text{nd}}$ is the probability that $E$ is detected due to imperfect cloning. Thus, we obtain $P_{A = B}^{\ket{\tilde{i}}} = p_1^{\ket{\tilde{i}}}+p_2^{\ket{\tilde{i}}}$. \textcolor{black}{The mutual information between $A$ and $B$, given that the prepared state is $\ket{\tilde{i}}$, is given by
\begin{equation}
    I_{AB}^{\ket{\tilde{i}}} =  \log_2 d - H(P_{A = B}^{\ket{\tilde{i}}}),
    \label{eq:I_AB_tilde}
\end{equation}
where $H(p)$ is the Shannon entropy of the probability distribution $\{p,(1-p)/(d-1),...,(1-p)/(d-1)\}$, since the $(d-1)$ possible errors are equiprobable. The average mutual information, over the $2d$ states, between the honest parties is 
\begin{eqnarray}
    I_{AB} &=&  \frac{1}{2d} \sum_i\left(I_{AB}^{\ket{i}} +  I_{AB}^{\ket{\tilde{i}}}\right)\nonumber\\
    &=&\log_2d-\frac{1}{2d}\sum_iH(P_{A = B}^{\ket{\tilde{i}}})\nonumber\\
    &=&\log_2d-\frac{1}{2}H\left(P_{A=B}^{\ket{\tilde{i}}}\right),
    \label{eq:I_AB}
\end{eqnarray}
}
where in the last line, we exploit that in the $F = F' = 1$ limit, the probability of Alice and Bob obtaining identical information, given that the input state is $\ket{\tilde{i}}$, is independent of the index $i$ (see Eq.~(\ref{eq:e-tilde_orthogonality})).
\subsubsection{Eve's mutual information with Alice and Bob}
\label{subsubsec:I_AE}
Given that the sets $\{ \ket{\epsilon_{ii}} \}$ and $\{ \ket{\eta_{ii}} \}$ comprise non-orthogonal states $E$ may misinterpret $\ket{\epsilon_{ii}} \ket{\eta_{jj}}$ for $\ket{\epsilon_{i'i'}} \ket{\eta_{j'j'}}$. This leads to the following three distinct situations concerning her ability to guess the correct ensemble:
\begin{enumerate}
    \item Both $\ket{\epsilon_{ii}} ~ \text{and} ~ \ket{\eta_{jj}}$ are identified correctly, in which case, the ensemble, and hence, the encoding operation, is guessed. This occurs with probability $p_\epsilon \times p_\eta$, where $p_{\epsilon (\eta)}$ is the probability of successful discrimination of the $d$ states $\ket{\epsilon_{ii}(\eta_{ii})}$.
    \item Either $\ket{\epsilon_{ii}} ~ \text{or} ~ \ket{\eta_{jj}}$ is misidentified, and according to Table.~\ref{tab:Eve_ensembles}, this leads to guessing the wrong ensemble and $E$ is unsuccessful in gaining any information about the message.
    \item Both states are identified incorrectly. In this situation, $E$ may still be successful in guessing the correct ensemble, e.g., if the state is $\ket{\epsilon_{00}} \ket{\eta_{11}}$ and she identifies $\ket{\epsilon_{00}}$ as $\ket{\epsilon_{11}}$ and $\ket{\eta_{11}}$ as $\ket{\eta_{22}}$, she correctly guesses the encoding operation as $\hat{U}_{11}$ despite being wrong on both occasions (see second column of Table.~\ref{tab:Eve_ensembles}). While misidentifying $\ket{\epsilon_{ii} (\eta_{ii})}$ occurs with a probability $(1 - p_{\epsilon (\eta)})$, the probability that two incorrect guesses leads to the correct ensemble is $(d - 1)^{-1}$. 
\end{enumerate}  
The total probability that $E$ can guess the actual encoding operation is thus

\begin{eqnarray}
    \hspace{-2em} P_{A=E}^{\ket{i(\tilde{i})}} =P_{A=E}= p_\epsilon p_\eta + \frac{(1 - p_\epsilon)(1 - p_\eta)}{(d - 1)}~~~\forall~\{\ket{i},\ket{\tilde{i}}\},\nonumber\\
    \label{eq:p_eve_guess}
\end{eqnarray}
\textcolor{black}{while the average mutual information over all input states, between Alice and Eve is 
\begin{equation}
    I_{AE} = \log_2 d - H(P_{A=E}).
    \label{eq:I_AE}
\end{equation}
}

Lastly, $I_{BE}$ comprises instances when the information with $B$ and $E$ is identical. The probability that this occurs is given by

\begin{equation}
    P_{B = E}^{\ket{i(\tilde{i})}} =  P^{\ket{i(\tilde{i})}}_{A = B} P^{\ket{i(\tilde{i})}}_{A = E} + \frac{1}{d - 1} P^{\ket{i(\tilde{i})}}_{A \neq B} P^{\ket{i(\tilde{i})}}_{A \neq E},
    \label{eq:p_B=E}
\end{equation}
where $P_{X \neq Y}^{\ket{x}} = 1 - P_{X = Y}^{\ket{x}}$ represents the probability that the dit values of $X$ and $Y$ differ when the input state is $\ket{x}$. Therefore, 
\begin{eqnarray}
    I_{BE}=\log_2d-\frac{1}{2d}\sum_i\left(H(P_{B=E}^{\ket{i}})+H(P_{B=E}^{\ket{\tilde{i}}})\right),
    \label{eq:I_BE}
\end{eqnarray}
and we are now equipped to compute the secret key rate, $r_d$, by comparing $I_{AB}$ with $I_{AE}$ and $I_{BE}$.

\subsubsection{Dimensional advantage on the key rate}
\label{subsubsec:cloning_results}

Our analysis so far has proceeded with the assumption that $E$ can clone the states $\{ \ket{i} \}$ perfectly, thereby making her detection probability minimum. As an exemplary case, let us further consider $\theta_{ij} = \theta ~ \text{and} ~ \theta'_{ij} = \theta' ~ \forall ~ i,j$, i.e., the pairwise overlap between all the states $\{ \ket{\epsilon_{ii}} \}$ and $\{ \ket{\eta_{ii}} \}$ is same. Moreover, $E$ can extract optimal information from both the channels when $\theta = \theta'$, otherwise, $E$ obtains more information through her attack on one channel than the other, which is not optimal (see Lemma in Ref.~\cite{Lucamracini_PRL_2005_LM05}). Using these conditions in Eq.~\eqref{eq:detection_prob}, the minimum probability of detecting $E$ reads as
\begin{eqnarray}
    \bar{P}_{\text{det}}^{\min} = \frac{d - 1}{d^2} \Big( (d + 1) + (d - 1) \cos \theta \Big) \sin^2 \frac{\theta}{2}.
    \label{eq:min_detection_prob}
\end{eqnarray}
Note that this also quantifies the amount of disturbance introduced by the eavesdropper, and hence is an estimate on how much information she can intercept. A protocol is more secure if it can exhibit a positive key rate for higher values of $\bar{P}_{\text{det}}^{\min}$. Furthermore, $\bar{P}_{\text{det}}^{\min}$ reaches its maximum at $\theta = \frac{\pi}{2}$, where it takes the value $\frac{d^2 - 1}{2d^2}$.

The aforementioned conditions also imply $P_{A=B}^{\ket{{i}}} = 1~\forall~i$ and $P_{A=B}^{\ket{\tilde{i}}} = \Big( 1 + (d - 1) \cos^2 \theta \Big)/d~\forall~i$, and hence
\begin{equation}
    I_{AB} = \log_2 d - \frac{1}{2} H \Big( \frac{1 + (d - 1) \cos^2 \theta}{d} \Big),
    \label{eq:I_AB_key-rate}
\end{equation}
while it is necessary to know $p_\epsilon$ and $p_\eta$ to gauge $I_{AE}$ and consequently, the final key rate.

Recall that $p_{\epsilon(\eta)}$ is the probability of discriminating the non-orthogonal states from the ensemble $\{ |\epsilon_{ii}(\eta_{ii})\rangle \langle \epsilon_{ii}(\eta_{ii})| \}$. The theory of minimum error discrimination~\cite{Helstrom_JSP_1969_min-error-discrimination, Holevo_JMA_1973_min-error-discrimination, Barnett_AOP_2009_state-discrimination-review, Bae_JPA_2015_state-discrimination-review} states that the maximum probability of successfully identifying one of the states is $p_{\epsilon(\eta)}^{\max} = \max_{\Pi_i} \frac{1}{d} \sum_{i = 0}^{d - 1}\Tr(\Pi_i |\epsilon_{ii}(\eta_{ii})\rangle \langle \epsilon_{ii}(\eta_{ii})| )$. While a multitude of numerical methods exist~\cite{Fiurasek_PRA_2002_numerical-min-error-discrimination, Eldar_IEEE_2003_SDP-state-discrimination, Wengang_IEEE_2008_SDP-min-error-discrimination} to calculate this optimal probability, let us assume that the optimal measurement is a set of projectors, $\{ \Pi_i\}$, such that $\Tr(\Pi_i |\epsilon_{ii}(\eta_{ii})\rangle \langle \epsilon_{ii}(\eta_{ii})| ) = \cos^2\chi~\forall~i$. This assumption of equiangular overlap may be justified by the fact that all the states are considered to have a pairwise overlap of $\theta$. For orthogonal ensembles, i.e., $\theta = \pi/2$, we must have $\cos^2\chi = 1$ since the optimal measurement reduces to the projectors of the states themselves. On the other hand, for $\theta = 0$, we should have $\cos^2\chi = 1/d$. Respecting these boundary conditions, we can infer that $\chi = \arccos \frac{1}{\sqrt{d}}(1 - 2 \theta/\pi)$. Under such conditions, $p_\epsilon = p_\eta = (1 + \cos 2 \chi)/2$, and \textcolor{black}{thus from Eq.~\eqref{eq:p_eve_guess} and (\ref{eq:I_AE}), we have}
\begin{equation}
    I_{AE} = \log_2 d - H \Big( \frac{d + \cos 2 \chi ( d \cos 2 \chi + 2d - 4)}{4 (d - 1)} \Big).
    \label{eq:I_AE_key-rate}
\end{equation}
Finally, using Eq. (\ref{eq:I_BE}), $I_{BE}$ in terms of $P_{A=B}^{\ket{\tilde{i}}}$ and $P_{A=E}$ (identifiable from Eqs.~\eqref{eq:I_AB_key-rate} and ~\eqref{eq:I_AE_key-rate}), can be written as 
\begin{eqnarray}
   \hspace{-2 em} I_{BE} &=& \log_2d-\frac{1}{2}H\left(P_{A=E}\right)\nonumber\\&& -\frac{1}{2}H \Big(P_{A=B}^{\ket{\tilde{i}}}P_{A=E}+\frac{1}{d-1}(1-P_{A=B}^{\ket{\tilde{i}}}) \times \nonumber \\ && ~~~~~~~~~~~~~~~~~~~~~~~~~~~~~~~~~~~~~~~~~~~~~~~\left(1-P_{A=E}\right)\Big).\nonumber\\
\end{eqnarray}

\begin{figure}[t]
\includegraphics[width=1.0\linewidth]{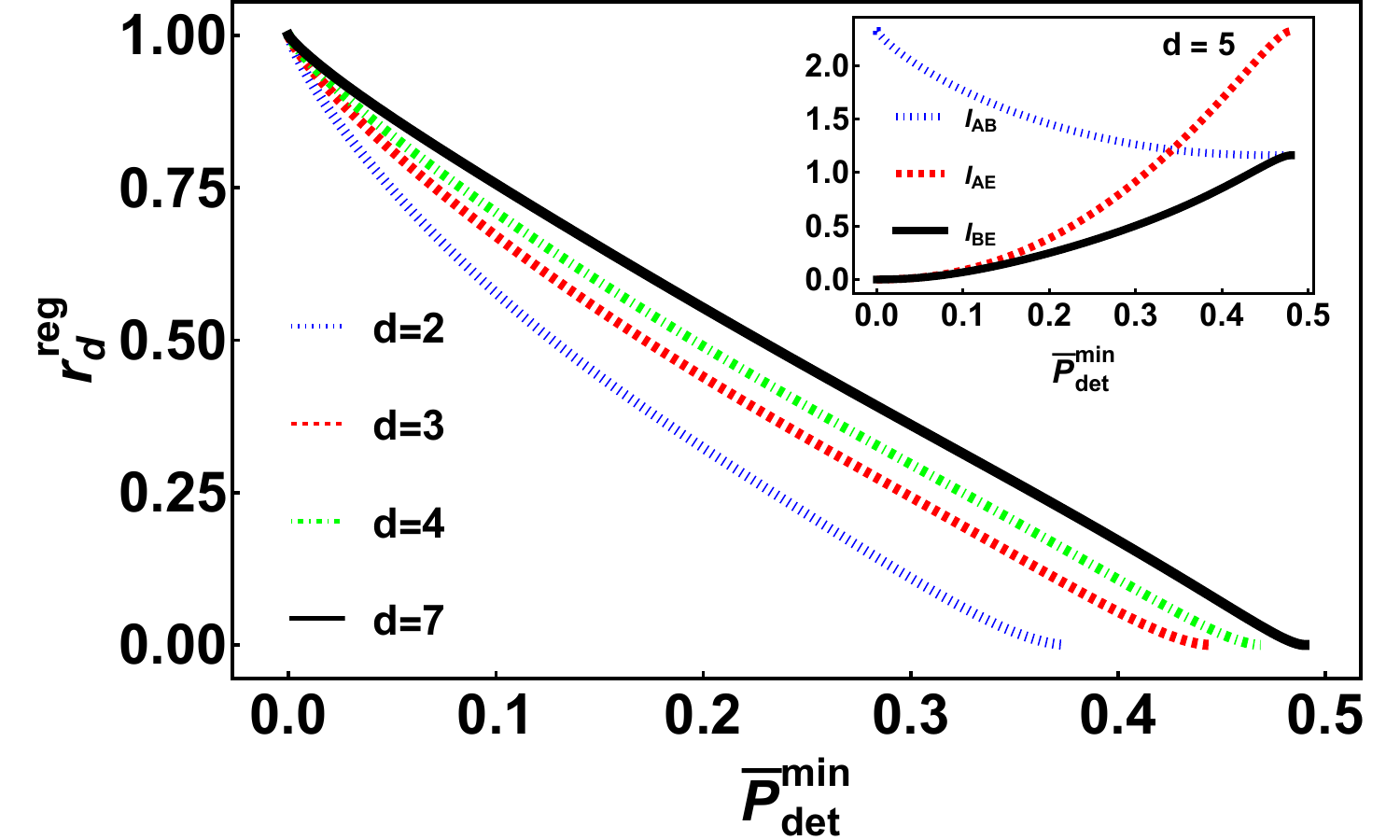}
\caption{\justifying {Lower bounds on the regularized key rate, $r_d^{\text{reg}}$ (ordinate), vs $\bar{P}_{\text{det}}^{\min}$ (abscissa) for two- (blue dotted), three- (red dashed), four- (green dot-dashed), and seven- (black solid) dimensional systems. The inset illustrates the variation of mutual Shannon information quantities, $I_{AB}$ (blue dotted), $I_{AE}$ (red dashed), and $I_{BE}$ (black solid) (ordinate) with respect to $\bar{P}_{\text{det}}^{\min}$ (abscissa) for $d = 5$. Both axes are dimensionless. 
}}
\label{fig:ind_attack1} 
\end{figure}

\begin{figure}[t]
\includegraphics[width=1.0\linewidth]{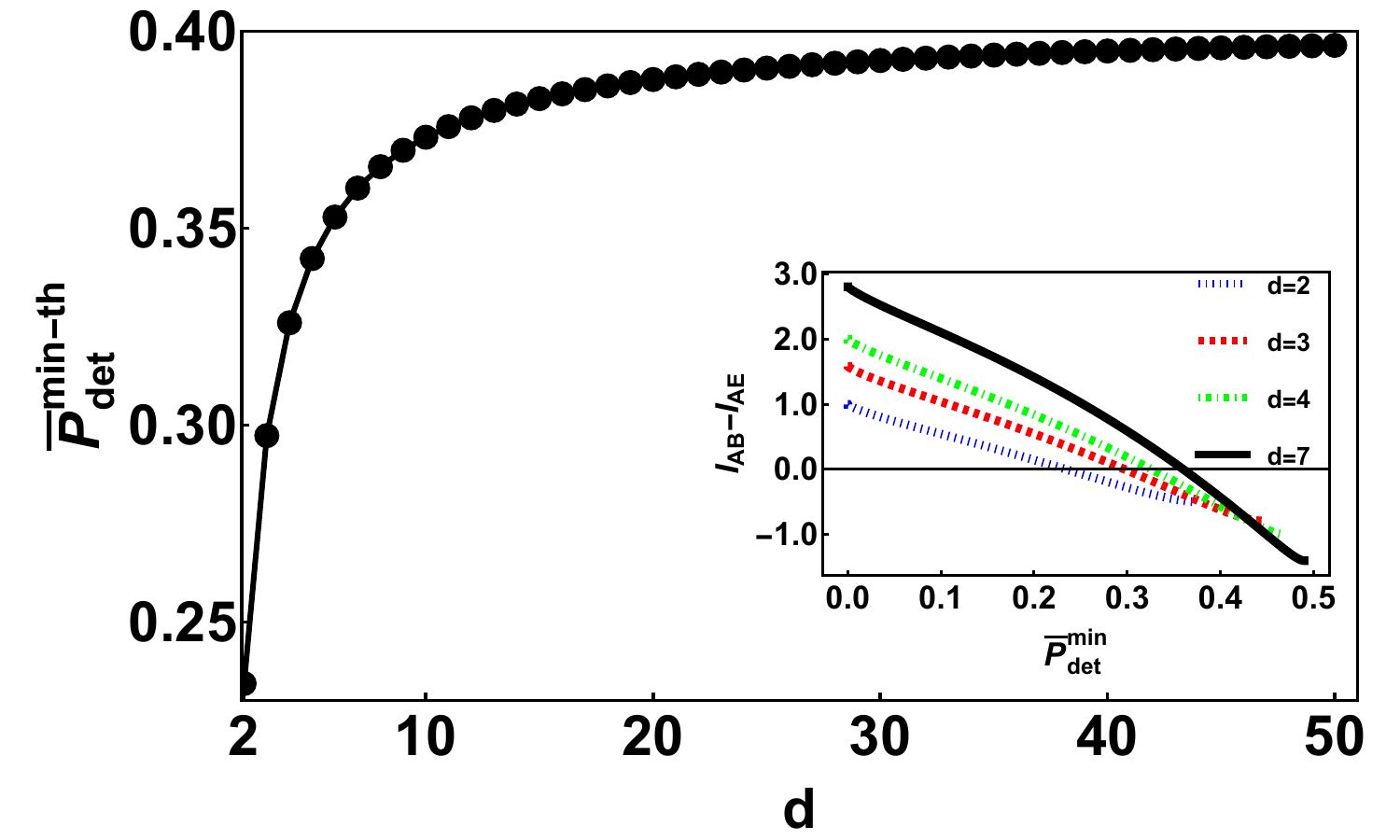}
\caption{\justifying {The threshold value of the minimum detection probability, $\bar{P}_{\text{det}}^{\min\text{-th}}$ (ordinate), against the dimension, $d$ (abscissa). (Inset) $I_{AB} - I_{AE}$ (ordinate) as a function of $\bar{P}_{\text{det}}^{\min}$ (abscissa) for two- (blue dotted), three- (red dashed), four- (green dot-dashed), and seven- (black solid)-dimensional systems. Both axes are dimensionless. }}
\label{fig:ind_attack2} 
\end{figure}

Thus, we have both the key rate and the minimum detection probability as functions of the parameter $\theta$, and we can track the variation of the former with increasing strengths of intervention. The lower bound on the regularized key rate, $r^{\text{reg}}_d = \frac{r_d}{\log_2 d}$, increases with an increase in the dimension for any specific value of the minimum detection probability of Eve, $\bar{P}_{\text{det}}^{\min}$ (see Fig.~\ref{fig:ind_attack1}). Furthermore, the range of $\bar{P}_{\text{det}}^{\min}$ permitting positive key rates expands with increasing dimension $d$. In particular, the maximum value of $\bar{P}_{\text{det}}^{\min}$ that still allows a positive key rate is $\frac{d^2 - 1}{2d^2}$, which increases monotonically with $d$. Thus, higher-dimensional systems provide a distinct advantage on two fronts - enhanced secret key rate as well as increased tolerance to third-party attacks. Further, the behavior of the mutual Shannon information quantities, $I_{AB}$, $I_{AE}$, and $I_{BE}$ possesses two notable features: $(i)$ Eve's mutual information with both Alice and Bob grows steadily with $\bar{P}_{\text{det}}^{\min}$, reaching their maximum values at $\bar{P}_{\text{det}}^{\min} = \frac{d^2 - 1}{2d^2}$, and $(ii)$ $I_{BE} \leq I_{AB}$ holds across all strengths of eavesdropping, guaranteeing that secret key distribution remains possible, as shown in the inset of Fig.~\ref{fig:ind_attack1}. These characteristic behaviors remain consistent across all dimensions. Moreover, as the dimension of the systems constituting the protocol increases, the threshold value of the minimum detection probability, $\bar{P}_{\text{det}}^{\min-\text{th}}$, beyond which $I_{AE}\geq I_{AB}$ also increases, and, \textcolor{black}{as a result, direct communication may become insecure, depending on the direction of classical communication during the post-processing stage}. This is further vindicated in Fig.~\ref{fig:ind_attack2} where we explicitly show how $\bar{P}_{\text{det}}^{\min-\text{th}}$ increases steadily with $d$. In other words, the protocol remains secure even though the eavesdropper can access more information about the exchanged messages, thereby showing higher resilience to adversarial attacks with increasing dimensions.

\section{Security analysis of the high-dimensional LM$05$ protocol under collective attack}
\label{sec:collective_attack}

Let us now consider the case when the eavesdropper has access to quantum memory and can hence perform joint measurements to obtain more information, popularly known as collective attack~\cite{Gisin_RMP_2002_cryptography-review}. In the regime of {\it collective attacks}, the eavesdropper resorts to appending a probe to each traveling qudit and measuring all the probes together after the honest parties have performed classical post-processing, which includes basis reconciliation, privacy amplification, and one-way error correction operations~\cite{Biham_PRL_1997_collective-attack-security, Gisin_RMP_2002_cryptography-review}.

\textcolor{black}{To coherently derive the secure key rate when the protocol suffers from collective attacks, we consider
a purification~\cite{Nielsen_2010_book} of the state-preparation and encoding operations~\cite{Beaudry_PRA_2013_two-way-QKD}. } This formalism assumes that the eavesdropper has control over the protocol from the very beginning, such that she/he prepares a tripartite state whereafter two parties are distributed to Alice and Bob. Unbeknownst to the honest parties, Eve can perform all possible allowed quantum operations to gather the maximum amount of information. Thus, it is crucial not to assume any knowledge of the devices involved to determine whether the protocol can furnish secure key strings. In this section, we assume that the devices of the two honest parties are characterized by a constant overlap factor~\cite{Koashi_NJP_2009_QKD-security-complimentarity} that is ultimately used together with entropic uncertainty relations~\cite{Koashi_JPCS_2006_QKD-entropic-uncertainty, Berta_NP_2010_undertainty-principle-memory, Coles_RMP_2017_entropic-uncertainty-relations-review} to arrive at the key rate. Note that the fundamental steps of the protocol are the same as delineated in the previous section, with certain modifications introduced to ensure heightened security against the more powerful class of attacks.

\subsection{Purified LM$05$ protocol}
\label{subsec:purified_LM05}

Two purification procedures are employed - one for preparing the states, and the other for the encoding operation. We again denote the participants as $A, B, ~\text{and}~ E$ for brevity.

\noindent \textbf{Purification of preparation}: In the purified preparation stage, $B$ initially measures one system of a maximally entangled state, $\ket{\phi^+}_{BB'} = \frac{1}{\sqrt{d}}\sum_j\ket{j,j}_{BB'}$, either in the basis $\{\ket{i}_{B'}\}$ or $\{\ket{\tilde{i}^c}_{B'} = \ket{\widetilde{d - i}}_{B'}\}$ to prepare $\ket{i}_{B}$ or $\ket{\tilde{i}}_{B}$ respectively, which is then sent to $A$, i.e.,
\begin{equation}
    \rho_A = \mathcal{E}_{1}^{B \to A}\Big( _{B'}\bra{i(\tilde{i}^c)} \rho^{\phi^+}_{BB'}\ket{i(\tilde{i}^c)}_{B'} \Big).
    \label{eq:purified_prep}
\end{equation}
The measurement outcome, $i$, is recorded as the preparation dit.

\noindent \textbf{Purification of encoding}:  The purified encoding operation, with probability $c \to 1$, is performed in such a way that the output is equivalent to that where $A$ uses a random string to decide which generalized Weyl operator is used (see Lemma $1$ in Ref.~\cite{Beaudry_PRA_2013_two-way-QKD}). This is achieved by appending a Bell state, $\rho^{\phi^+}_{A''A'}$, to the original qudit, $\rho_A$, and measuring in the $d$-dimensional Bell basis, $\ket{B(xy)}_{AA''} = \frac{1}{\sqrt{d}}\sum_{l=0}^{d-1} \omega^{-l~y} \ket{l,l\oplus x}$~\cite{Bertlmann_JPA_2008_qudit-Bloch-vector}, with $x,y = 0, 1, \dots, d - 1$. Specifically~\cite{Patra_PRA_2024_dimensional-advantage-SDC}, 
\begin{eqnarray}
    \hspace{-2.5em}    \rho'^{xy}_{A'} &=& \hat{U}_{xy}^{A\to A'} \rho_A \hat{U}_{xy}^{\dagger A\to A'} \nonumber \\
        &=& d^2 \times ~_{AA''}\bra{B(xy)} \rho_A \otimes \rho^{\phi^+}_{A''A'}\ket{B(xy)}_{AA''}.~~~
    \label{eq:purified_encoding}
\end{eqnarray}
In this process the output of the aforementioned encoding operation is independent of the input state, when averaged over all encodings, i.e., $\frac{1}{d^2}\sum_{xy}\rho'^{xy}_{A'}=\frac{\mathbb{I}}{d}$.
After the encoding operation, the encoded state is sent back to $B$, through the channel $\mathcal{E}_2^{A' \to B}$. The choice of encoding is recorded by $A$ as a two-dit string $(x,y)$.


\noindent \textbf{Purified version of the check-run}: When $A$ decides to switch to check-mode, with probability $(1 - c)$, she measures the received qudit in the computational or the Fourier basis. Upon getting outcome $\ket{i(\tilde{i})}$, she measures one half of a Bell state in the basis $\{\ket{i(\tilde{i}^c)}\}$ to send $\ket{i(\tilde{i})}$ to $B$ through $\mathcal{E}_2^{A' \to B}$, i.e., we have $\mathcal{E}_2^{A' \to B} \Big( 
 {_A}\bra{i(\tilde{i})} \rho_A \ket{i(\tilde{i})}{_A}~ _{A''}\bra{i(\tilde{i}^c)} \rho^{\phi^+}_{A'' A'} \ket{i(\tilde{i}^c)}_{A''}  \Big)$.

\noindent \textbf{Decoding the message}: Similar to the original protocol, the decoding measurement by $B$ is in the preparation basis, $\{\ket{i}\}$ or $\{\ket{\tilde{i}}\}$. Assume that the measurement yields the outcome $j$. After repeating the steps $N$ times, the parties perform classical post-processing wherein $B$ reveals which basis the initial states were prepared in, while $A$ declares which qudits were encoded and which were measured. In our work, we concentrate on reverse basis-reconciliation~\cite{Beaudry_PRA_2013_two-way-QKD}, and $A$ also discloses the basis in which the check measurements were performed. The key with $B$ is then the sum of his preparation dit $i$ and the complement of the measurement outcome $j^c=(d-j)$, modulo $d$, while $A$ retains the first recorded dit, $x$, if the initial state was prepared in the computational basis or the second dit, $y$, if it was prepared in the Fourier basis.

\textcolor{black}{\textbf{Note $\mathbf{2}$:} While the original protocol involves encoding through $\hat{U}_{xx}$ alone (see Sec.~\ref{sec:individual_attack}), the purified encoding scheme produces additional states $\rho'^{xy}_A ~\text{with}~ x \neq y$. The modified LM$05$ protocol reduces to the original one once Bob reveals his preparation basis. However, in the original protocol, although Bob never discloses his choice of basis, Alice's encoding operations are also restricted to $\hat{U}_{xx}$ instead of $\hat{U}_{xy}$.}

\subsubsection*{Coarse-graining of encoding and measurements} 

The aforementioned description of the purified protocol involves modulo-$d$ addition of dits (by $B$) and choosing one out of the two recorded encoding dits (by $A$) in order to generate the secret key shared by the honest parties. The necessity of this step stems from the fact that states other than those encoded through $\hat{U}_{xx}$ are generated due to the completeness of the measurement basis, $\{\ket{B(xy)}\}$, in the encoding scheme. Incorporating this modification in the analysis comprises the use of rank-$d$ projectors as the encoding, decoding, and check measurements. Considering, $\hat{M}(i,j) = \ket{ij}\bra{ij}$ and $\hat{B}(ij)=\ket{B(ij)}\bra{B(ij)}$ with $\{i,j\}\in\{0,1,\cdots,d-1\}$, we enumerate the coarse-grained encoding and decoding measurements as well as the corresponding check measurements, as follows:  
\begin{enumerate}
    \item \textbf{Computational basis runs:} \begin{eqnarray}
     \hspace{-3.5em}  && \nonumber \begin{rcases}
    \hspace{-3.5em}     \text{encoding: }\mathbb{B}_0(x)_{AA''}=\sum_y \hat{B}(xy)_{AA''} \\
    \hspace{-3.5em}       \text{decoding: }\mathbb{M}_0(x)_{BB'}=\sum_k \hat{M}(k\oplus x,k)_{BB'}
       \end{rcases} ~ \text{message}, \\
        \label{eq:coarse_comp-encoding} \\
       && \begin{rcases}
    \hspace{-3.5em}   \mathbb{N}_0(x)_{AA''} = \mathbb{I}_A\otimes \ket{\tilde{x^c}}\bra{\tilde{x^c}}_{A''} \\
    \hspace{-3.5em}   \mathbb{N}_0(x)_{BB'} = \ket{\tilde{x}}\bra{\tilde{x}}_B\otimes \mathbb{I}_{B'}
       \end{rcases} ~\text{check} \label{eq:coarse_comp-testB}
    \end{eqnarray}
    \item \textbf{Fourier basis runs:} 
    \begin{eqnarray}
       && \nonumber \begin{rcases}
       \hspace{-3.5em}    \text{encoding: }\mathbb{B}_1(y)_{AA''}=\sum_x \hat{B}(xy)_{AA''} \\
       \hspace{-3.5em}    \text{decoding: }\mathbb{M}_1(y)_{BB'}=\sum_k \hat{M}(\widetilde{k\oplus y},\tilde{k^c})_{BB'}
       \end{rcases} ~ \text{message}, \\
        \label{eq:coarse_fourier-encoding} \\
       && \begin{rcases}
     \hspace{-3.5em}  \mathbb{N}_1(x)_{AA''} = \mathbb{I}_A\otimes \ket{x}\bra{x}_{A''} \\
    \hspace{-3.5em}   \mathbb{N}_1(x)_{BB'} = \ket{x}\bra{x}_B\otimes \mathbb{I}_{B'} \label{eq:coarse_comp-testB}
       \end{rcases} ~\text{check} \label{eq:coarse_fourier-test}
    \end{eqnarray}
\end{enumerate}
  \label{eq:key-run_purify}

\subsection{Recipe to compute the key rate}
\label{subsec:modified_key-rate}

\textcolor{black}{When performing a collective attack, the eavesdropper is assumed to possess a global pure state, $\ket{\psi}_{A A'' B B' E}$~\cite{Beaudry_PRA_2013_two-way-QKD}, parts of which are distributed to the honest parties. As $A$ and $B$ carry out the protocol, $E$ can measure her/his corresponding subsystem to obtain information about the secret message. 
As there are two different preparation bases at the disposal for $B$, denoted by $\theta = 0,1$, corresponding to the computational and Fourier bases respectively, we obtain the final state by averaging over the preparations which occur with a probability, $p_\theta$, as}
\begin{eqnarray}
  \hspace{-4.5em} \nonumber &\kappa_{A A'' B B' E}& \\ 
   \nonumber \hspace{-4.5em}&=& \sum_{\theta=0}^1\sum_{x,x' =0}^{d-1} p_{\theta} \Big[ \mathbb{B}_\theta(x)_{AA''} \otimes \mathbb{M}_\theta(x')_{BB'} \Big] \Big(\rho^{\psi}_{AA''BB'E}\Big)\\
    \hspace{-3.5em} &=&\sum_{\theta=0}^1\sum_{x,x' =0}^{d-1} p_{\theta}q_{xx'}^{\theta} \ket{x}\bra{x}_A\otimes \ket{x'}\bra{x'}_B\otimes\rho_{E}^{xx'\theta}\nonumber\\
    \hspace{-3.5em} &=& \sum_{\theta=0}^1 p_{\theta} \kappa^{\theta}_{AA''BB'E},
     ~~~ \label{eq:kappa_state}
\end{eqnarray}
where
$$\kappa^{\theta}_{AA''BB'E}=\sum_{x,x' =0}^{d-1} \Big[ \mathbb{B}_\theta(x)_{AA''} \otimes \mathbb{M}_\theta(x')_{BB'} \Big] \Big(\rho^{\psi}_{AA''BB'E}\Big).$$ 
Note that $\ket{x}_A$ and $\ket{x'}_B$ are classical registers recording the raw keys, and $\rho_{E}^{xx'\theta}$ is the state at $E$ depending on the measurement statistics, $q^{\theta}_{x x'}$, enabling the eavesdropper to gather information about the protocol.

To derive the key rate, let us define two fictitious states 
\begin{eqnarray}
  \hspace{-4.5em} \nonumber &\tau_{A A'' B B' E}& \\
  \nonumber \hspace{-4.5em}  &=& \sum_{\theta=0}^1\sum_{x,x' =0}^{d-1} p_{\theta}\Big[ \mathbb{N}_\theta(x)_{AA''}\otimes \mathbb{M}_\theta(x')_{BB'} \Big] \Big(\rho^{\psi}_{AA''BB'E}\Big), \\
 \hspace{-4.5em}  &=&\sum_{\theta=0}^1 p_{\theta}\tau_{A A'' B B' E}^{\theta} \label{eq:tau_state}  \nonumber 
 ~,~\text{and}~~
\end{eqnarray}
\begin{eqnarray} 
    \nonumber \hspace{-4.5em} & \sigma_{A A'' B B' E}& \\
   \nonumber \hspace{-4.5em} &=& \sum_{\theta=0}^1\sum_{x,x' =0}^{d-1} p_{\theta}\Big[ \mathbb{N}_\theta(x)_{AA''}\otimes \mathbb{N}_\theta(x')_{BB'} \Big] \Big(\rho^{\psi}_{AA''BB'E}\Big). \\
  \hspace{-4.5em}  &=& \sum_{\theta=0}^1 p_{\theta}\sigma_{A A'' B B' E}^{\theta}.
   \label{eq:sigma_state}
\end{eqnarray}

The lower bound on the key rate, $r_d$, is an average rate over two preparation bases, i.e., $r_d = \sum_{\theta=0}^1 p_{\theta}r_d^{\theta}$, where, $r_d^{\theta}$ represents the lower bound on the key rate corresponding to preparation in the $\theta$ basis  
\begin{eqnarray}
   r_d^{\theta} &=& I_{AB,\Theta=\theta}^{\kappa} - I_{BE,\Theta=\theta}^{\kappa} \label{eq:key-rate_1}\\
    &=& S_{B|E,\Theta=\theta}^{\kappa} - S_{B|A,\Theta=\theta}^{\kappa} \label{eq:key-rate_2} \\
    &=& \textcolor{black}{S_{B|E,\Theta=\theta}^{\tau} - S_{B|A,\Theta=\theta}^{\kappa} \label{eq:key-rate_3}} \\
    &\geq& \log_2 \frac{1}{\gamma} - S_{B|A,\Theta=\theta}^{\sigma} - S_{B|A,\Theta=\theta}^{\kappa} \label{eq:key-rate_4}.
\end{eqnarray}
\textcolor{black}{Here, the subscript $\Theta = \theta$ with $\theta = 0,1$ denoting the preparation in the computational and Fourier bases respectively}. In this protocol, $p_\theta$ is assumed to be $1/2$ for both preparations. In Eqs.~(\ref{eq:key-rate_1})-(\ref{eq:key-rate_4}), the superscript over the mutual information and von Neumann entropy terms, $S_A^\sigma \equiv S(\sigma_A) = -\Tr (\sigma_A \log_2 \sigma_A)$~\cite{Nielsen_2010_book}, denote the state for which they are evaluated and the definition of the mutual information, $I_{AB} = S_B - S_{B|A}$, in terms of the conditional entropy, $S_{B|A} = S_{AB} - S_{A}$, is used to proceed to Eq.~\eqref{eq:key-rate_2}. From the expressions of $\kappa_{A A'' B B' E}$ and the fictitious state $\tau_{A A'' B B' E}$, it is evident that $S_{B|E}^{\kappa} = S_{B|E}^{\tau}$ as the only difference in these states is the measurements at $A$ which do not affect the conditional entropy between $B$ and $E$, whence Eq.~\eqref{eq:key-rate_3} follows. Lastly, we use the entropic uncertainty relation\footnote{Corresponding to two different POVM settings, denoted as $P_X\equiv\{P^i_X\}_i$ and $P_Z \equiv\{P^j_Z\}_j$, with classical outcomes, $i$ and $j$ respectively, performed by the party A on a tripartite state $\rho_{ABE}$, the entropic uncertainty relation states 
\begin{equation}
    S_{X|E}+S_{Z|B}\geq \log_2 \frac{1}{\gamma},
\end{equation}
where $\gamma = \max_{i,j} ||\sqrt{P^i_X}\sqrt{P^j_Z}||_\infty^2$~\cite{Berta_NP_2010_undertainty-principle-memory, Coles_RMP_2017_entropic-uncertainty-relations-review}.}, $S_{B|E}^{\sigma} + S_{B|E}^{\tau} \geq \log_2 1/\gamma$ with $\gamma = \max_{x, x'} ||\sqrt{\hat{\mathcal{M}_\theta}(x)} \sqrt{\hat{\mathcal{N}_\theta}(x')}||_{\infty}^2$ \textcolor{black}{(for two POVMs $\{\hat{\mathcal{M}_\theta}(x)\}_x$ and $\{\hat{\mathcal{N}_\theta}(x')\}_{x'}$)} to derive the final expression for the lower bound on the key rate, \textcolor{black}{ using the Schatten $\infty$-norm, $||\cdot||_{\infty}$, which is defined as the largest singular value of the operator.}

\subsection{Key rate of LM$05$ under collective attack}
\label{subsec:collective_results}

We now present our results on the behavior of the secure key rate of the LM$05$ protocol when subjected to collective attacks. Given that the discussion so far is based on the eavesdropper preparing a global pure state before supplying the required resource, the honest parties always begin the protocol with a mixed state. We can, therefore, analyse the protocol's performance by considering that it is affected by environmental noise, which can be due to the presence of the eavesdropper. The noise models we consider include the dit-phase flip and depolarizing channels, which are examples of Pauli channels, and the non-unital $d$-dimensional amplitude damping channel (ADC). This section is devoted to studying the variation of the key rate against increasing noise strengths and demonstrating how employing qudits provides more resilience to external influences and leads to a higher key rate than the protocol with qubits. 

Furthermore, we also compare the LM$05$ protocol with another well-known two-way quantum key distribution scheme known as {\it secure dense coding} (SDC)~\cite{Bostrom_PRL_2002_ping-pong, Han_SR_2014_ping-pong-security, Beaudry_PRA_2013_two-way-QKD}, which has already been shown to offer increased quantum advantage when using higher-dimensional resources~\cite{Patra_PRA_2024_dimensional-advantage-SDC}. Unlike the LM$05$ protocol, which can generate at most one secure dit per run, SDC can be used to securely transmit more than one dit, in particular, two dits of information per run in a noiseless scenario due to the presence of pre-shared entanglement in a two-qudit system. We, therefore, investigate whether a $d^2$-dimensional LM$05$ ($d^2-$LM$05$), can match a $d$-dimensional SDC ($d-$SDC), which employs a $d \otimes d$-dimensional entangled state, or whether two parallel $d$-dimensional LM$05$s $\Big(2\times(d-\text{LM}05) \Big)$ are sufficient for the task. Note that all these protocols produce the same key rate in the absence of noise.

We commence our analysis by examining the statistics obtained during the message and check phases, assuming the two communication channels (forward and backward) to be independent, \textcolor{black}{ while the case for correlated channels is deferred to a later section.} Two distinct nontrivial probabilities can be identified --- $q^\theta_{x = x'}$ and $q^\theta_{x \neq x'}$ for the message step, and $\tilde{q}^\theta_{x = x'}$ and $\tilde{q}^\theta_{x \neq x'}$ for the check runs. These represent the instances of $A$ and $B$ obtaining either the outcome with $x = x'$ or the outcome with $x \neq x'$, respectively. They are, moreover, dependent on each other. \textcolor{black}{For Pauli channels, which affect all states of a given generalized Weyl eigenbasis equally, this simplifies to $d q^\theta_{x=x'} + d(d-1) q^\theta_{x\neq x'}=1$ (and similarly for $\tilde{q}^\theta$) since differing outcome probabilities occur $d(d - 1)$ times in the joint probability distribution, while matching outcome probabilities occur only $d$ times.} 

\textcolor{black}{The total probability of Alice and Bob receiving different outcomes is referred here as the \textit{ error rate} and is represented as $Q$. For the message run, it is generally termed as the \textit{quantum dit error rate} (QDER)}. In this analysis, we refer to the error rate corresponding to the message run and the check run by $Q_k$ and $Q_t$, respectively. 
\\
\textbf{1. Independent Pauli Noise: }
For independent Pauli channels, $Q_k=d(d-1)  q^{\theta}_{x\neq x'}$ and $Q_t$ can be obtained by replacing $q^\theta_{x\neq x'}$ with $\tilde{q}^{\theta}_{x\neq x'}$. In terms of the error rate, the conditional entropy reads

\begin{eqnarray}
    S_{B|A}^{\kappa(\sigma)} &=& S_{AB}\Big |_{\rho=\kappa(\sigma)} - S_A\Big |_{\rho=\kappa(\sigma)}\nonumber\\
    &=&  -[(1-Q_{k(t)})\log_2(1-Q_{k(t)}) \nonumber\\&& \hspace{2.2 cm}+ Q_{k(t)}\log_2\frac{Q_{k(t)}}{(d-1)}].
    \label{eq:conditionalentropy}
\end{eqnarray}
The depolarizing channel furnishes the input state, $\rho$, admixed with white noise, i.e., $\mathcal{E}_{\text{dep}}^{B\to A}(\rho_B)=(1-p)\rho_A+\frac{p}{d}\mathbb{I}_{d_A}$ (where $\mathbb{I}_{d_A}$ is the $d$-dimensional identity in the Hilbert space of $A$) whereas, under the dit-phase flip channel, $\mathcal{E}_{\text{dpf}}^{B\to A}(\rho_B) = (1-p)\rho_A+\frac{p}{(d-1)^2}\sum_{i,j =1}^{d-1} \hat{W}_{ij}^{B\to A}\rho_B \hat{W}_{ij}^{\dagger B\to A}$, with $p$ denoting the noise strength, \textcolor{black}{and $\hat{W}_{ij} = \sum_{k = 0}^{d-1} e^{\frac{2 \pi \iota}{d} k i} |k \rangle \langle k\oplus j|$~\cite{Fonseca_PRA_2019_high-dim-teleportation-noisy, Lidar_arXiv_2019_open-quantum-systems-notes}.} 
\textcolor{black}{When the forward and backward channels are independent, i.e., $\rho \to \mathcal{E}_2^{A\to B} \circ \mathcal{U}^A \circ \mathcal{E}_1^{B\to A} (\rho)$ (here, $\mathcal{U}^A$ denotes the channel corresponding to encoding by $A$), and identical, with the same noise strength, $p$, the error rates can be expressed as
\begin{eqnarray}
    Q_k &=& 
    \begin{cases}
        \frac{d-1}{d} p(2-p) & \text{for } \mathcal{E}_{\text{dep}},\\
        \frac{1}{d-1} p[2(d-1)-dp] & \text{for } \mathcal{E}_{\text{dpf}},
        \label{indep. depo. Q}
    \end{cases}\\
    \text{and}\,\, Q_t &=&
    \begin{cases}
        \frac{d-1}{d} p & \qquad \qquad \qquad~~\text{for } \mathcal{E}_{\text{dep}},\\
        p & \qquad \qquad \qquad~~\text{for } \mathcal{E}_{\text{dpf}}.
        \label{indep. dpf. Q}
    \end{cases}
\end{eqnarray}}
\noindent \textcolor{black}{Note that the depolarizing channel and the dit-phase flip channel treat both the computational and Fourier basis states symmetrically. As a result, the error rate and consequently the lower bound on the key rate, remain the same regardless of the basis from which the initial states are sampled. Hence, Eqs.~(\ref{indep. depo. Q}) and~(\ref{indep. dpf. Q}) do not have any $\theta$-dependence, which leads to 
$r_d^0=r_d^1$. However, this does not hold for all Pauli channels. For instance, while the dephasing channel preserves computational basis states perfectly, it introduces rotations when acting on Fourier basis states~\cite{Lidar_arXiv_2019_open-quantum-systems-notes}, demonstrating basis-dependence.} The marginal probability distribution for Pauli channels generated at Alice's part is always uniform and equal to $\frac{1}{d}$, and thus, $S_A = \log_2 d$. It can be shown that in this scenario, but not in general, the key rates corresponding to preparation in different bases during the message run are equal, enabling us to consider any one case. The key rates for the depolarizing and the dit-phase flip channels can be obtained from Eq.~\eqref{eq:key-rate_4} by substituting the expressions of the probability functions in Eqs.~\eqref{indep. depo. Q} and ~\eqref{indep. dpf. Q} in Eq.~\eqref{eq:conditionalentropy} while \textcolor{black}{noting that $\gamma = \max_{x, x'} ||\sqrt{\hat{\mathcal{M}_\theta}(x)} \sqrt{\hat{\mathcal{N}_\theta}(x')}||^2 = 1/d$.} 


\begin{figure}[t]
\includegraphics[width=1.0\linewidth]{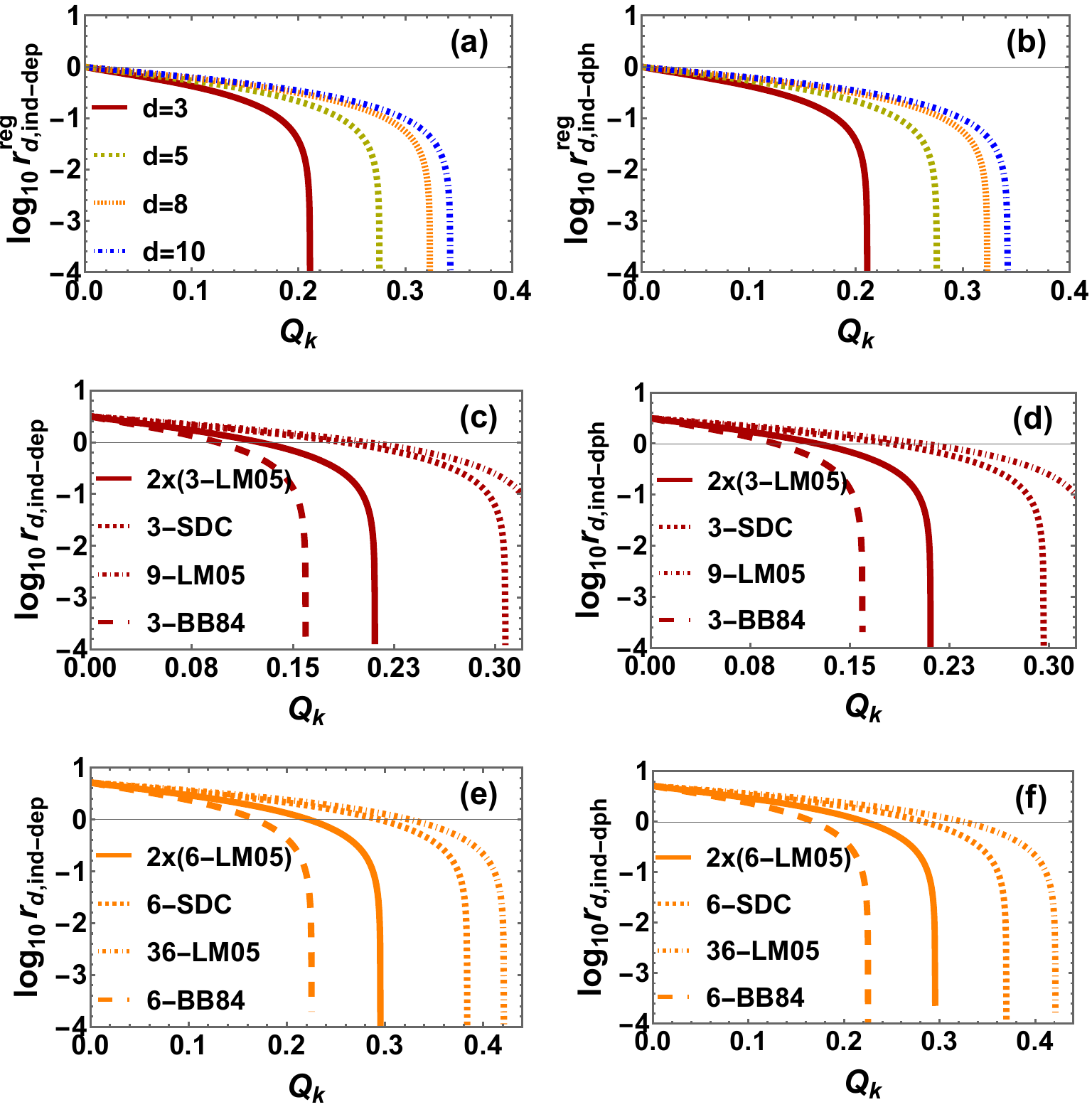}
\caption{\justifying {Lower bounds on the key rate (ordinate) for independent depolarizing and dit-phase flip channels as functions of respective QDERs, $Q_k$ (abscissa). Upper panel: Dimensional analysis using $r^{\text{reg}}_d$, for LM$05$ with $d = 3$ (red solid), $d = 5$ (yellow dashed), $d = 8$ (orange dotted), and $d = 10$(blue dash-dotted), for depolarizing $(a)$ and dit-phase flip $(b)$ noises. Middle panel: $(c)$ and $(d)$ compare $2 \times ($d$-$LM$05)$ (solid), $d^2-$LM$05$ (dash-dotted), $d-$SDC (dashed) and $2 \times (d-$BB$84)$ (long dashed) protocols with $d = 3$ (red).
Lower Panel: $(e)$ and $(f)$ compare the same protocols as the middle panel with $d = 6$ (orange) for depolarizing (left panels) and dit-phase flip channels (right panels), respectively. Both axes are dimensionless.}}
\label{fig:coll_attack_ind} 
\end{figure}
\begin{figure}[t]
\includegraphics[width=1.0\linewidth]{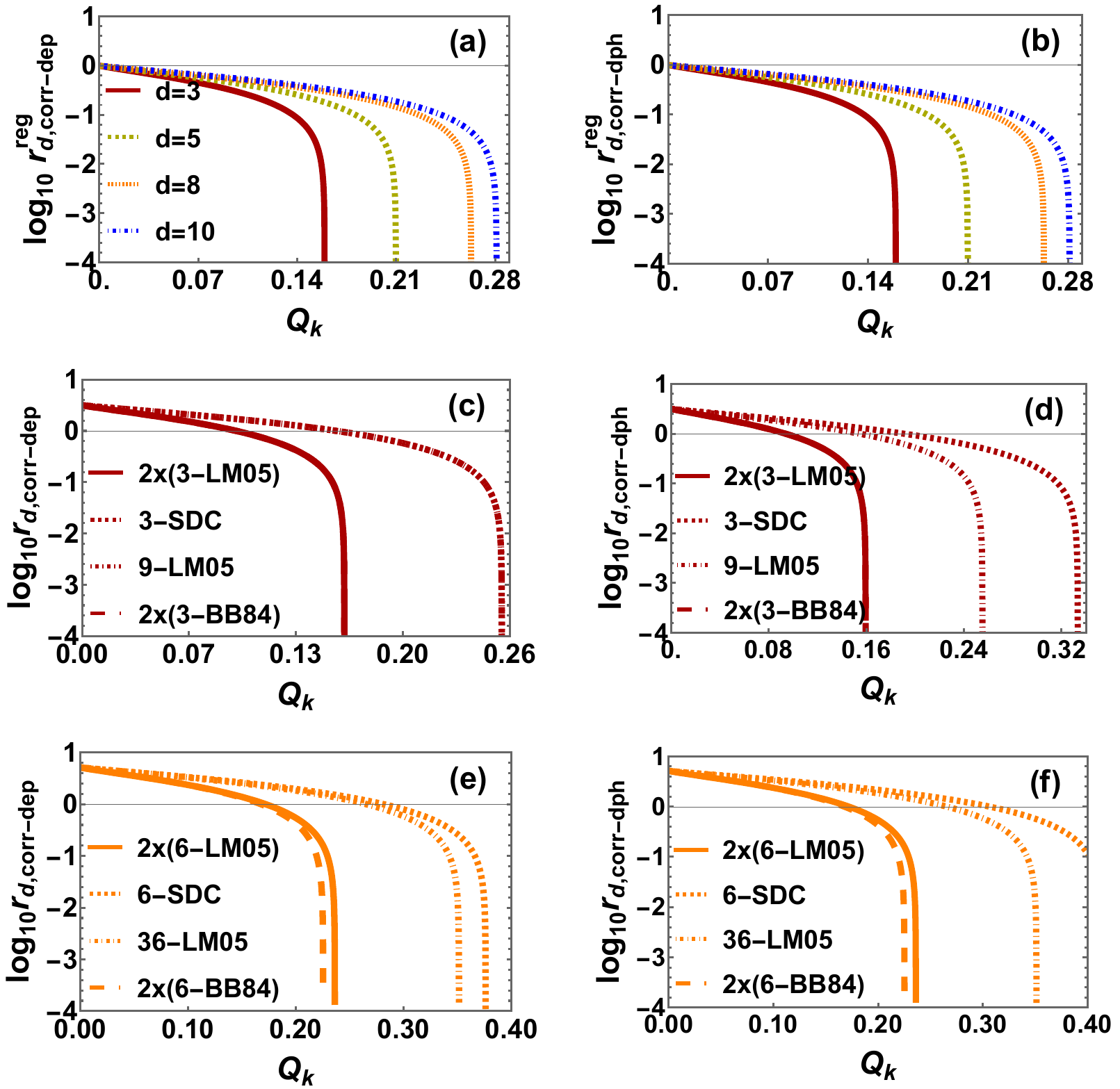}
\caption{\justifying {Lower bounds on the key rate (ordinate) for correlated depolarizing and dit-phase flip channels with respect to the QDER, $Q_k$, (abscissa). All other specifications are the same as in Fig.~\ref{fig:coll_attack_ind}}.}
\label{fig:coll_attack_corr} 
\end{figure}

\textcolor{black}{We observe that the regularized key rate, $r_d^{\text{reg}}$, increases with dimension $d$ at a fixed QDER, $Q_k$, indicating a dimensional advantage (see Figs.~\ref{fig:coll_attack_ind} (a) and (b)).} Moreover, one obtains substantial improvement in key rates when using a $d^2$-dimensional LM$05$ protocol as compared to two parallel $d$-dimensional LM$05$ protocols, as shown in Figs.~\ref{fig:coll_attack_ind} (c) to (f). This is evident, as an eavesdropper can interfere only twice in the former, while, in the latter, the opportunities for eavesdropping are effectively doubled. Intriguingly, the comparison of a $d^2$-dimensional LM$05$ with the existing $d$-dimensional SDC, in Figs.~\ref{fig:coll_attack_ind} (c) to (f), demonstrates that when independent Pauli noises are considered, the LM$05$ protocol has an edge over SDC. This can be attributed to the difference in noise dimensionality — while the noise in SDC operates in $d$ dimensions,  it acts on the full $d^2$-dimensional space in the $d^2$-dimensional LM$05$ scheme. \textcolor{black}{For completeness, we also include a comparison with the one-way BB$84$ scheme \cite{Bennett_TCS_2014_BB84}. To ensure a fair comparison with the other two-way protocols discussed here, we consider two copies of one-way BB$84$ protocols: one from Bob to Alice and the other from Alice to Bob (henceforth, two-way BB$84$). This configuration yields two secure key dits, in analogy with the two parallel implementations of the $d$-dimensional LM$05$ protocol, a single $d$-dimensional SDC protocol, or a single $d^2$-dimensional LM$05$ protocol. It is worth noting that, in the presence of independent noise, the two-way LM$05$ protocol outperforms the two-way BB$84$ (see Fig.~\ref{fig:coll_attack_ind}).}  In terms of the key rate, we thus encounter the following hierarchy for the independent Pauli channels under study: 
\begin{center}
    \begin{equation}
       2 \times (d-\text{BB}84) < 2 \times (d-\text{ LM}05) < d-\text{SDC} < d^2-\text{LM}05.
        \label{eq:hierarchy}
    \end{equation}
\end{center}
\textbf{$\mathbf{1 (a).}$ Correlated Pauli channels:} Another realization of Pauli noise involves correlation between the forward and backward channels~\cite{Bowen_PRA_2004_memory-channel, Giovannetti_JPA_2005_memory-channel, Kretschmann_PRA_2005_memory-channels, Bayat_PRA_2008_memory-spin-channel, Caruso_PRA_2008_qubit-memory-channel, Caruso_RMP_2014_memory-channel-review}. Physically, this may correspond to the situation when the same channel is used in quick succession such that its properties are unaltered~\cite{Macchiavello_PRA_2002_memory-channel-information}. The evolution of the state when such perfect correlation is present may be represented as
\begin{eqnarray}
    \hspace{-3.0em}&& \rho_B \xrightarrow[]{\mathcal{E}^{B \to A}_{\text{corr-dep}}} \Big(1 - \frac{d^2 - 1}{d^2} p\Big)  \rho_B \nonumber \\
  \hspace{-3.0em} && + \frac{p}{d^2} \sum_{\substack{i,j = 0 \\ i + j \neq 0}}^{d - 1} \left(\hat{W}_{ij}^{A\to B} (\cdot)\hat{W}_{ij}^{B\to A}\right) \rho_B \left(\hat{W}_{ij}^{A\to B} (\cdot)\hat{W}_{ij}^{B\to A}\right)^\dagger \label{eq:depo_corr}, \nonumber\\
 \hspace{-3.0em}  &&\text{and}\nonumber\\
  \hspace{-3.0em} && \rho_B \xrightarrow[]{\mathcal{E}^{B \to A}_{\text{corr-dpf}}} (1 - p) \rho_B  \nonumber \\
  \hspace{-3.0em} && + \frac{p}{(d - 1)^2} \sum_{i,j = 1}^{d - 1} \left(\hat{W}_{ij}^{A\to B} (\cdot)\hat{W}_{ij}^{B\to A}\right) \rho_B \left(\hat{W}_{ij}^{A\to B} (\cdot)\hat{W}_{ij}^{B\to A}\right)^\dagger,\nonumber  \label{eq:deph_corr}
\end{eqnarray}
where \((\cdot)\) denotes the encoding operation. \textcolor{black}{Here we explicitly consider that the channels are fully correlated, i.e., a perfect correlation between the channels exist.} \textcolor{black}{Note that the action of the same unitaries, $\hat{W}_{ij}$, before (forward channel) and after (backward channel) the encoding operation
is the fundamental difference with the uncorrelated noise model, $\mathcal{E}_2 \circ \mathcal{U}^A \circ \mathcal{E}_1 (\rho) = \sum_{i,j,k,l} \hat{W}_{kl} (\cdot) (\hat{W}_{ij} \rho \hat{W}_{ij}^\dagger) (\cdot) \hat{W}_{kl}^{\dagger}$.} As a consequence, the destructive effect of the forward channel can be mitigated by the backward one~\cite{Lucamarini_TCS_2014_QKD-two-way-channel}, as in the case of polarisation drift in fiber optics~\cite{Marand_OL_1995_30km-QKD, Yuan_OE_2005_QKD-fibre, Dixon_APL_2010_high-bit-rate-QKD-polarization}. Therefore, it is interesting to study whether correlations between the two noisy channels yield any advantage over the key rate obtained in the presence of independent channels. Note that correlated channels have been shown to yield better communication capacities~\cite{Macchiavello_PRA_2004_channel-capacity-two-qubit-memory, Bowen_PRA_2005_classical-capacity-memory-channel, Arshed_PRA_2006_correlated-channel-classical-capacity, Das_PRA_2014_noise-inverts-dense-coding} and teleportation fidelities~\cite{Li_IJTP_2019_correlated-channel-teleportation}, which have also been demonstrated experimentally~\cite{Ball_PRA_2004_communication-correlated-birefringence, Banaszek_PRL_2004_correlation-communication-experiment}.

For the correlated depolarizing and dit-phase flip channels, the error rates relevant for the key rate analysis are given by
\begin{eqnarray}
    Q_k &=& 
    \begin{cases}
        \frac{f(d)-1}{f(d)} p & \text{for } \mathcal{E}_{\text{dep}},\\
         \frac{(f(d)-1)/f(d)}{(d-1)/d}p & \text{for } \mathcal{E}_{\text{dpf}},
        \label{eq:corr_Q_k}
    \end{cases}
\end{eqnarray}
where \textcolor{black}{$f(d):=\frac{d}{2}\Big( 1 + (d \mod{2}) \Big)$}, while $Q_t$ follows the same structure, with $f(d)$ being replaced by $d$. The conditional entropy terms \( S_{B|A}^{\kappa(\sigma)} \) can now be obtained from Eq.~\eqref{eq:conditionalentropy}, with \( d \) replaced by \( f(d) \) in the case \textcolor{black}{of \( S_{B|A}^{\kappa} \),} using the error rates \( Q_k \) and \( Q_t \) provided above. 

\begin{figure}[t]
\begin{center}
    \includegraphics[width=0.82\linewidth]{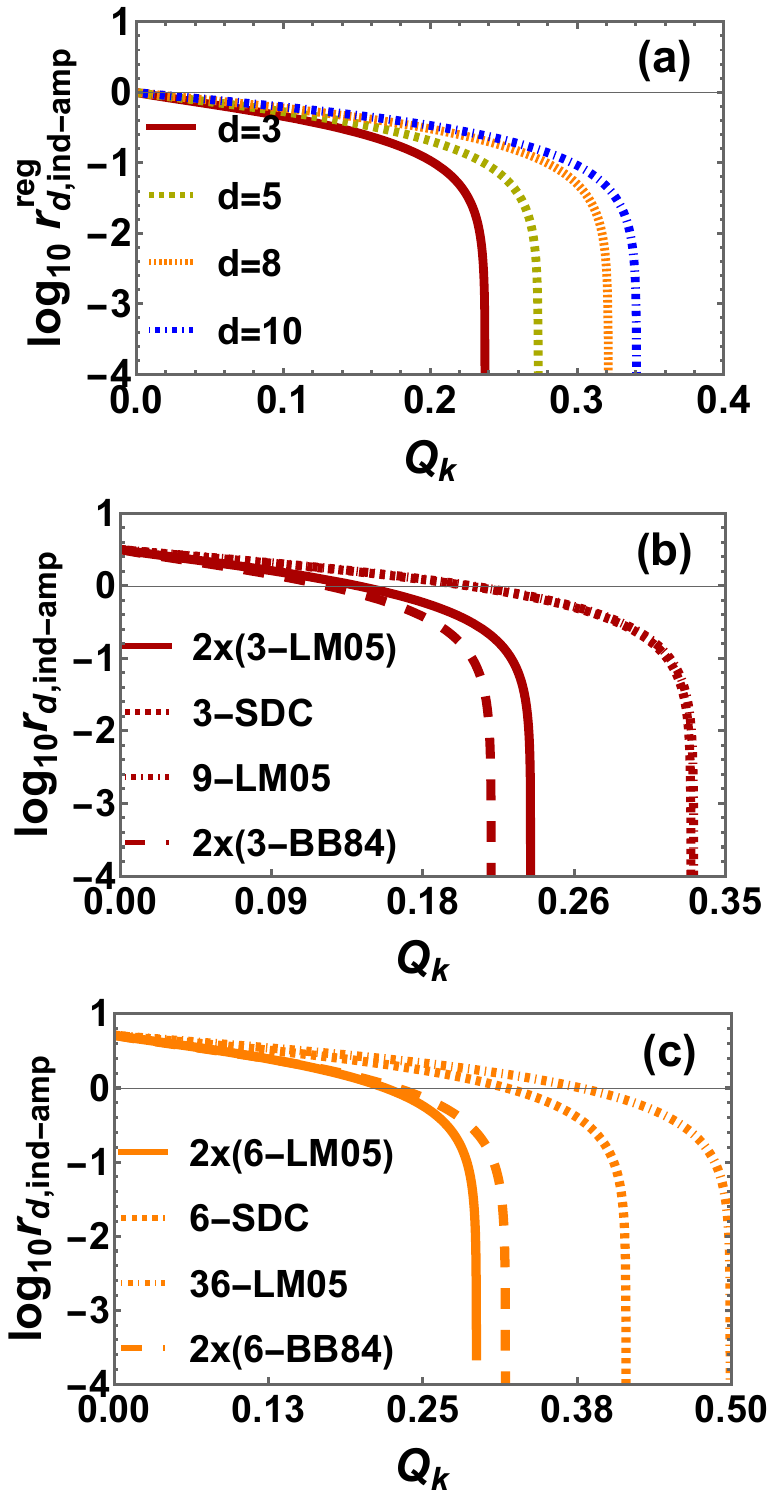}
\end{center}
\caption{\justifying {Lower bounds on the key rate (ordinate) for independent amplitude damping channels against the QDER, $Q_k$ (abscissa). Specifications for $(a)$, $(b)$, and $(c)$ are respectively the same as in the upper and lower panels of Fig.~\ref{fig:coll_attack_ind}}.}
\label{fig:coll_attackAmp} 
\end{figure}
\textcolor{black}{Due to the perfect correlation between the forward and backward channels, the correlated noise mitigates the bit-mismatch between $A$ and $B$, leading to a lower QDER. Thus, for the same QDER, the actual noise strength for correlated channels is much higher than that for the independent ones. As a result, the regularised key rate under correlated noise is lower than that under the corresponding independent model for the same QDER (compare upper panels of  Figs.~\ref{fig:coll_attack_ind} and~\ref{fig:coll_attack_corr}).}
Furthermore, under the correlated noise model, the hierarchy observed in the independent noise scenario changes. We observe that the $d-$SDC protocol consistently outperforms the entanglement-free $d^2-$LM$05$ one under correlated noise, except for odd dimensions where $d^2-\text{LM}05$ matches the performance of $d-$SDC under depolarizing noise (see Figs.~\ref{fig:coll_attack_corr} (c) to (f)). These findings suggest that the presence of entanglement may enhance the robustness of the protocol against correlated noise. \textcolor{black}{Another noteworthy aspect of the analysis concerns the comparison between two realisations of a one-way protocol, such as BB$84$, and a genuine two-way protocol, such as LM$05$, in the presence of correlated noise. As shown in Figs.~\ref{fig:coll_attack_corr} (c) to (f), two parallel implementations of LM$05$ outperform the two-way BB$84$ only for even dimensions. In contrast, for odd dimensions, the corresponding key rates coincide exactly.}

\textbf{Note $\mathbf{3}$:} From Eq.~(\ref{eq:corr_Q_k}), it is evident that for \( d = 2 \), the quantum bit error rate (QBER), \( Q_k \), vanishes despite Eve's intervention. However, the key rate still decreases as Eve's interference increases, which can be detected through the error rate observed in the check runs. This observation highlights that a zero QBER does not necessarily guarantee the security of the protocol.

\textbf{$\mathbf{2}$. Non-unital noise:} While the Pauli noises are examples of unital channels, which leave the maximally mixed state invariant~\cite{Nielsen_2010_book}, the presence of non-unital noise makes the hierarchy of the concerned key distribution protocols less strict. A notable example of a non-unital noise model is the amplitude-damping channel (ADC)~\cite{Lidar_arXiv_2019_open-quantum-systems-notes}, used to model the spontaneous decay of photons from an excited state to the ground state~\cite{deOliveira_PLA_2020_generalised-amplitude-damping-channel}, characterized by the  Kraus operators~\cite{Fonseca_PRA_2019_high-dim-teleportation-noisy},
\begin{eqnarray}
  \hspace{-1.7em}  \nonumber K_0 = \ket{0}\bra{0}+\sqrt{1-p}\sum_{i=1}^{d-1}\ket{i}\bra{i} ~ \text{and} ~ K_i = \sqrt{p}\ket{0}\bra{i}, \\ \label{eq:amp_damp_Kraus}
\end{eqnarray}
with $i = 1, \dots, d - 1$.
For independent ADCs modelling the forward and backward transmissions, we derive the expressions for the error rates as
\begin{eqnarray}
   \hspace{-1.5em} Q_k = 
    \begin{cases}\nonumber
        \frac{d-1}{d}p(2-p) \qquad \qquad \qquad \qquad\qquad\quad \text{ for }\theta=0,\\
        \frac{d-1}{d^3} \Big(4 (d - 2) (1 - \sqrt{1 - p}) + 
   p \Big((2 d (d - 2)  + 4)\\  - (d - 2)^2 p- 4 (d - 2) (1 - \sqrt{1 - p})\Big) \Big)\\
~~~~~~~~~~~~~~~~~~~~~~~~~~~~~~~~~~~~~~~~~~~~~~~~~~~~~~~~\text{ for}\,\theta=1,\\
    \end{cases}
    \label{eq:amp_Q_k}
\end{eqnarray}
and
\begin{eqnarray}
  \hspace{-1.5em}  Q_t = 
    \begin{cases}
        \frac{d-1}{d^2}\Big(2-2\sqrt{1-p}+(d-2)p\Big) & \text{for }\theta =0,\\
        \frac{d-1}{d} p & \text{for }\theta =1.
    \end{cases}
\end{eqnarray}
The conditional entropy corresponding to the state $\kappa_{AA''BB'E}$ is obtained from Eq.~(\ref{eq:conditionalentropy}), using the error rate $Q_k$ defined in Eq.~(\ref{eq:amp_Q_k}). \textcolor{black}{Contrary to Pauli channels, however, the number of non-trivial differing outcome probabilities for $\theta = 1$ in the case of $\sigma_{AA''BB'E}$ is only $d-1$. 
The expression of conditional entropy corresponding to \( \sigma_{AA''BB'E} \) becomes
\begin{eqnarray}
  \hspace{-1.5em}  S_{B|A}^{\sigma}=
    \begin{cases}
        -\Big(Q_t\log_2Q_t-(\frac{d-1}{d}-Q_t)\log_2(\frac{d-1}{d}-Q_t)\Big)\\
      \qquad+ \frac{d-1}{d} \log_2 \frac{d-1}{d} \qquad\qquad \text{for }\theta=1,\\
        -\Big((1-Q_t)\log_2(1-Q_t)+ Q_t\log_2\frac{Q_t}{(d-1)}\Big).\\ \qquad \qquad \qquad \qquad \qquad\qquad  \text{for }\theta=0.
    \end{cases}
\end{eqnarray}}
The key rates depend on the preparation basis adopted during the message run, and thus, the overall key rate is given by the average over those corresponding to the computational and Fourier bases. Fig.~\ref{fig:coll_attackAmp} shows the result of our study highlighting the effect of the amplitude damping channel on the lower bound of the secret key rate. Fig.~\ref{fig:coll_attackAmp} (a) demonstrates that, in the case of independent ADC, the regularized key rate increases with the dimension, hinting at the universality of dimensional advantage for generic quantum channels. Moreover, though the hierarchy of efficiency of the three considered protocols, $2 \times (d-\text{ LM}05), d-\text{SDC}, ~\text{and}~ d^2-\text{LM}05$, remain unchanged for even dimensions, as in the case of the independent Pauli channel, it changes when the dimensions are odd. Specifically, $d-$SDC and $d^2-$ LM$05$ are equally efficient as can be seen in Fig.~\ref{fig:coll_attackAmp} (b) for $d=3$. \textcolor{black}{Another striking difference that can be observed in the case of even dimensions is the reversal of order when we compare the two-way $d$-dimensional BB$84$ to two parallel implementations of $d$-dimensional LM$05$. The order is restored in the odd dimensions (see Fig.~\ref{fig:coll_attackAmp}(b) and (c)).} 


\section{Conclusion}
\label{sec:conclu}

The BB$84$~\cite{Bennett_TCS_2014_BB84} protocol is the earliest and most extensively studied quantum key distribution (QKD) scheme which  was proven to be secure against collective attacks~\cite{Lo_Science_1999_unconditional-security, Shor_PRL_2000_device-dependent-BB84, Mayers_JACM_2001_unconditional-security} up to a certain quantum bit error rate (QBER). Over time, it has undergone various modifications --  one such scheme is the \tm{entanglement-free two way quantum key distribution protocol, where the classical post-processing scheme is one way}~\cite{Lucamracini_PRL_2005_LM05}. It enables deterministic key generation by eliminating the need for sifting of keys. Originally designed for qubit systems, it has been demonstrated to be secure against the most general form of eavesdropping attacks~\cite{Lu_PRA_2011_LM05-secure, Fung_PRA_2012_LM05-secure, Beaudry_PRA_2013_two-way-QKD}. Although some research has been conducted on the security analysis of high-dimensional key distribution protocols, studies specifically addressing deterministic protocols remain scarce. Given the significant studies conducted towards high-dimensional implementation of information-theoretic protocols and the practical advantages of secure deterministic communication, it is imperative to address this gap.

In this work, we proposed the \tm{entanglement-free, two-way QKD} protocol using higher-dimensional systems and established its security against both individual cloning attacks, which do not rely on quantum memory, as well as against the most general class of attacks, namely collective attacks. Since the protocol involves the transmission of the constituent system twice, the eavesdropper, Eve, has the opportunity to attack both the forward and backward channels, thereby gaining more information at the cost of a higher probability of detection compared to one-way protocols. We derived a lower bound on the achievable key rate under both individual and collective attacks.

To assess the security of the protocol under individual attacks, we considered a general cloning operation, subject to certain simplifying assumptions, which is performed by Eve \tm{to minimize her detection probability, as well as gain the maximal possible information about the system of the honest parties}. The lower bound on the key rate, when varied against the minimum probability of detecting \tm{the presence of} Eve, revealed three distinct advantages with increasing dimension: $(i)$ for a fixed detection probability, higher-dimensional systems yield higher key rates, $(ii)$ the range of detection probabilities that allow for a positive key rate increases with dimension, and $(iii)$ the threshold value of the minimum detection probability beyond which the mutual information between the honest parties is lower than that with the eavesdropper, also increases with dimension. \tm{The last point makes our protocol more robust and secure, as it causes the eavesdropper to gain less information about the honest parties, while leaving more detectable traces of her presence.} Furthermore, under our assumptions, we demonstrated that the key distribution is possible at a finite rate, regardless of channel noise, for any dimension.

In the case of collective attacks, where the eavesdropper is more powerful owing to their access to quantum memory, we adopted an approach that lies between fully device-dependent and device-independent frameworks to analyze the performance of the protocol. We considered a purified model of the initial state preparation and encoding operations, allowing the malicious party to distribute the shared resource among the honest users, and assumed that they can perform any operation allowed by quantum mechanics. This enabled us to account for the most general class of eavesdropping attacks.
Further, in practice, all quantum information processing tasks are affected by noise. Due to imperfect isolation, quantum systems inevitably interact with their environment, which degrades their inherent quantum properties and consequently reduces protocol performance. In fact, in the collective attack scenario, noise can be used to model the presence of Eve. To analyze this problem, we examined three canonical noise models, the depolarizing, the dit-phase flip, and the amplitude-damping channels, and evaluated their impact on the secret key rate
with the corresponding quantum dit error rate (QDER). 

In the regime of collective attacks involving both individual and correlated noise, two distinct dimensional advantages emerge: $(i)$ for a fixed QDER, the key rate increases with increasing dimension; and $(ii)$ the range of detection probabilities that allow for a positive key rate increases with dimension. We also compare \tm{the higher dimensional entanglement-free two way} protocol with another well-known \tm{entanglement-based} two-way QKD scheme known as secure dense coding (SDC), and with the one-way BB$84$ protocol in a higher-dimensional scenario. For a fair comparison, we consider four scenarios, LM$05$ implemented using \( d^2 \)-dimensional systems (\( d^2 \)-LM$05$), SDC using \( d \otimes d \) entangled system (\( d \)-SDC), two parallel \( d \)-dimensional LM$05$ protocols (\( 2 \times (d-\text{LM}05) \)), and two copies of \( d \)-dimensional one-way BB$84$ (\( 2 \times (d-\text{BB}84) \)). While all four protocols achieve the same key rate in the absence of noise, their performances differ under noisy conditions. Specifically, under an independent noise model, the \( d^2- \)LM$05$ protocol outperforms the others, followed by \( d -\)SDC and then the \( 2 \times (d-\text{LM}05) \). On the other hand, under the correlated noise model, we observe that the \( d -\)SDC protocol outperforms the entanglement-free \( d^2 \)-LM$05$, except in the case of odd dimensions, where the performance of \( d^2 \)-dimensional LM$05$ matches with that of \( d -\)SDC under depolarizing noise. This indicates that entanglement enhances the protocol's resilience to correlated noise.

\textcolor{black}{The comparison between one-way and genuine two-way protocols is likewise instructive. We observe that two parallel implementations of the two-way LM$05$ protocol outperform two copies of the one-way BB$84$ protocol in nearly all cases when the applied noise is characterized by Pauli operators, except non-unital channels in even dimensions. Furthermore, this advantage is reinforced in the presence of correlated Pauli noise models for even dimensions; however, the advantage is negligible and disappears when the dimension is odd.}

\textcolor{black}{Interestingly, one can increase the number of bases to $d+1$ MUB-s when $d$ is a prime number, as it indeed offers significant advantages, lowering the probability that the eavesdropper correctly guesses the preparation basis and ensuring that any interception introduces more disturbance on average. However, this improvement comes with an inherent trade-off. Although the extra basis increases the sensitivity to eavesdropping, it reduces the raw key rate, since the probability that Alice and Bob choose the same basis decreases, leading to additional discarding of raw keys during basis reconciliation. For example, a comparison between the performance of the qudit-based one-way quantum key distribution using two MUBs (similar to BB$84$) and that using all $d+1$ of them has shown that the latter protocol yields a substantially lower key-rate while offering only a slight improvement in the tolerance to third-party interception~\cite{Cerf_PRL_2002_high-d-BB84-six-state}. A similar situation arises when comparing the BB$84$ protocol with the six state protocol, where the number of MUBs is increased from two to three. In this case, the QBER rises from $25\%$ to $33\%$ upon adding the third basis ~\cite{Bennett_TCS_2014_BB84, Bruss_PRL_1998_six-state, Bechmann_PRA_1999_six-state} for an intercept-resend attack. The additional basis enhances the symmetry of the protocol and strengthens the information–disturbance tradeoff, thereby improving the sensitivity to the intervention of a malicious third party, at the expense of a lower key rate. We expect a similar scenario to arise when increasing the number of MUB-s from two to $d+1$. On a related note, a protocol similar to LM$05$ has also been studied in prime power dimensions ($d = p^m$, $p$ being a prime number and $m$ being a positive integer) using only $d$ number of MUBs for encoding~\cite{Eusebi_QIC_2009_high-d-LM05}. In this case, as the dimension increases, the detection probability of Eve in each successful test run (i.e., when Alice and Bob choose the same basis) increases as $\frac{(d-1)^2}{d^2}$. However, the probability that the legitimate users select the same basis decreases as $\frac{1}{d}$. Consequently, the overall detection probability becomes $\frac{(d-1)^2}{d^3}$, which attains its maximum at $d=3$. Importantly, the analysis is restricted to a specific class of attacks in which the adversary applies a control-shift unitary operation. In contrast, our work adopts a more general framework, as we do not impose any restriction on the form of the attack and instead consider a general unitary operation in our analysis. Exploring this direction in greater depth is a promising avenue for future study.}



In summary, we have generalized the two-way deterministic quantum key distribution from qubit-based systems to protocols where the key is encoded in quantum states belonging to a Hilbert space of arbitrary finite dimensions. While the advantages of higher-dimensional systems in QKD are well-established, our results demonstrate their enhanced utility in deterministic two-way cryptographic schemes, further reinforcing the importance of larger Hilbert spaces in enhancing protocol performance.\\

\section*{Acknowledgment}

This research was supported in part by the ``INFOSYS Scholarship for senior students''. A.M. acknowledges funding support for Chanakya - PhD fellowship from the National Mission on Interdisciplinary Cyber Physical Systems, of the Department of Science and Technology, Govt. of India through the I-HUB Quantum Technology Foundation. R.G. acknowledges funding from the HORIZON-EIC-$2022$-PATHFINDERCHALLENGES-$01$ program under Grant Agreement No.~$10111489$ (Veriqub). Views and opinions expressed are however, those of the authors only and do not necessarily reflect those of the European Union. Neither the European Union nor the granting authority can be held responsible for them.

\appendix

\bibliographystyle{elsarticle-num-names}
\bibliography{ref}

\end{document}